\documentclass[conference]{IEEEtran}

\usepackage{cite}
\usepackage{amsmath,amssymb,amsfonts}
\usepackage{algorithmic}
\usepackage{graphicx,color}
\usepackage{textcomp}
\usepackage{xcolor}
\usepackage{algorithm,algorithmic}
\def\BibTeX{{\rm B\kern-.05em{\sc i\kern-.025em b}\kern-.08em
T\kern-.1667em\lower.7ex\hbox{E}\kern-.125emX}}
\AtBeginDocument{\definecolor{tmlcncolor}{cmyk}{0.93,0.59,0.15,0.02}\definecolor{NavyBlue}{RGB}{0,86,125}}

\usepackage{flushend}

\usepackage{amsmath}
\DeclareMathOperator*{\argmax}{arg\,max}

\usepackage{glossaries}
\newacronym{awgn}{AWGN}{additive white Gaussian noise}
\newacronym{iid}{i.i.d.}{independent and identically distributed}
\newacronym{snr}{SNR}{signal-to-noise ratio}
\newacronym{bcr}{BCR}{bandwidth compression ratio}
\newacronym{psnr}{PSNR}{peak signal-to-noise ratio}
\newacronym{mse}{MSE}{mean-squared error}
\newacronym{dl}{DL}{deep learning}
\newacronym{dnn}{DNN}{deep neural network}
\newacronym{deepjscc}{DeepJSCC}{deep joint source-channel coding}
\newacronym{jscc}{JSCC}{joint source-channel coding}
\newacronym{bsc}{BSC}{binary symmetric channel}
\newacronym{vimco}{VIMCO}{variational inference for Monte Carlo objectives}
\newacronym{ml}{ML}{maximum likelihood}
\newacronym{gdn}{GDN}{generalized divisive normalization}
\newacronym{ber}{BER}{bit error rate}
\newacronym{bler}{BLER}{block error rate}
\newacronym{ldpc}{LDPC}{low density parity check}
\newacronym{uep}{UEP}{unequal error protection}
\newacronym{csi}{CSI}{channel state information}

\usepackage{epstopdf}

\usepackage{pgfplots}
\usepackage{pgfplotstable}
\pgfplotsset{compat=1.17}
\usepackage[caption=false,font=footnotesize]{subfig}

\definecolor{DarkGreen}{rgb}{0.1,0.5,0.1}

\begin{document}

\newtheorem{theorem}{Theorem}
\newtheorem{lemma}{Lemma}
\newtheorem{definition}{Definition}
\newtheorem{corollary}{Corollary}
\newtheorem{example}{Example}
\newtheorem{claim}{Claim}
\newtheorem{assum}{Assumption}
\newtheorem{remark}{Remark}

\markboth{Multi-level reliability interface for semantic communications over wireless networks}{Tung {et al.}}

\title{Multi-level Reliability Interface for Semantic Communications over Wireless Networks}

\author{Tze-Yang Tung, Homa Esfahanizadeh, Jinfeng Du, and Harish Viswanathan\\
Nokia Bell Labs, Murray Hill, NJ 07974, USA\\
emails: \{tze-yang.tung, homa.esfahanizadeh, jinfeng.du, harish.viswanathan\}@nokia-bell-labs.com \vspace{-0.5cm}}

\maketitle

\begin{abstract}
    Semantic communication, when examined through the lens of joint source-channel coding (JSCC), maps source messages directly into channel input symbols, where the measure of success is defined by end-to-end distortion rather than traditional metrics such as block error rate. Previous studies have shown significant improvements achieved through deep learning (DL)-driven JSCC compared to traditional separate source and channel coding. However, JSCC is impractical in existing communication networks, where application and network providers are typically different entities connected over general-purpose TCP/IP links. In this paper, we propose designing the source and channel mappings separately and sequentially via a novel multi-level reliability interface. This conceptual interface enables semi-JSCC at both the learned source and channel mappers and achieves many of the gains observed in existing DL-based JSCC work (which would require a fully joint design between the application and the network), such as lower end-to-end distortion and graceful degradation of distortion with channel quality. We believe this work represents an important step towards realizing semantic communications in wireless networks.
\end{abstract}



\section{Introduction}
\label{sec:intro}

Semantic communication has received significant interest in recent years \cite{gündüz23_timel_massiv_commun, luo22_seman_commun, lu23_rethin_moder_commun_seman_codin_seman_commun, yang23_seman_commun_futur_inter, gunduz23_beyon_trans_bits}.
The concept originated from Shannon and Weaver's seminal work \cite{shannon_weaver_semantic}, which focuses on accurately conveying the ``meaning'' of the source message rather than  precisely reconstructing transmitted bits. 
The majority of published work on semantic communication propose to realize this concept through \gls{jscc}. \gls{jscc} integrates source coding with channel coding, directly mapping each source message to a channel codeword.
This approach contrasts with traditional communication system design, which has historically employed separate source and channel coding steps.
Theoretically, it has been shown that \gls{jscc} can achieve lower end-to-end distortion than separate source and channel coding in the finite blocklength regime  \cite{kostina13_lossy_joint_sourc_chann_codin}, which is relevant to practical communication systems.

To date, the majority of work approaching semantic communications through \gls{jscc} typically employs an autoencoder architecture with an untrainable channel between the encoder and decoder. The encoder learns a latent representation of the source that is resilient to channel distortions.
This technique has been successfully  applied and evaluated across various domains such as image \cite{bourtsoulatze19_deep_joint_sourc_chann_codin}, video \cite{tung22_deepw}, text \cite{xie21_deep_learn_enabl_seman_commun_system,Esfahanizadeh_ISIT_TexShape}, generative applications \cite{erdemir23_gener_joint_sourc_chann_codin}, and image classification \cite{jankowski21_wirel_image_retriev_edge}, and it has demonstrated security against eavesdroppers \cite{tung23_deep_joint_sourc_chann_encry_codin, kalkhoran23_secur_deep_jscc_again_multip_eaves,Esfahanizadeh_ICASSP_InfoShape}.
Furthermore, challenging channel conditions, such as multiple access channels, can be addressed this way using state-of-the-art \gls{dl} architectures \cite{yilmaz23_distr_deep_joint_sourc_chann,wu23_vision_trans_adapt_image_trans_mimo_chann}. 

While \gls{dl}-driven \gls{jscc} has shown promise, its implementation faces practical challenges within general-purpose TCP/IP communication networks. These networks are fundamentally designed to accommodate traffic from diverse sources through a modular architecture, where application and network providers are typically distinct entities.
Under this current framework, applications often reside in cloud data centers and communicate with end devices (i.e., application consumers) over multiple wireline hops, with the final hop being wireless. 
This wireless link typically comprises either a mobile cellular connection between a base station and the device or a WiFi link between an access point and the device. 
The application compresses the source data using a chosen codec, packaging it into blocks of bits that are then packetized and transmitted over the network. Conventionally, wireline hops over optical fiber links are treated as lossless, and the last wireless hop is viewed as the bottleneck link requiring optimization. 
The primary objective is typically to maximize the utilization of the wireless link, assuming unlimited resources for the wireline links. Consequently, \gls{jscc} focuses on optimizing for the wireless channel.

Another important challenge of employing \gls{jscc} over today's networks is the impracticality of sending real-time Channel State Information (CSI), such as \gls{snr}, of the wireless channel back to the source. In fact, base stations or access points, serving as intermediate network nodes, may not even be aware of the application's source address. Directly sending source data to the base station for \gls{jscc} processing is also infeasible. This would necessitate the base station becoming a destination node for the source, terminating the TCP/IP connection, and then creating a new connection to the device with the base station as the source node. 
Apart from lacking scalability, this approach requires the source application to disclose its data to the base station.
As a result, tackling semantic communications involving \gls{jscc} over current communication networks is  a challenging problem that we address in this paper.

\begin{figure}
    \centering
    \includegraphics[width=0.9\linewidth]{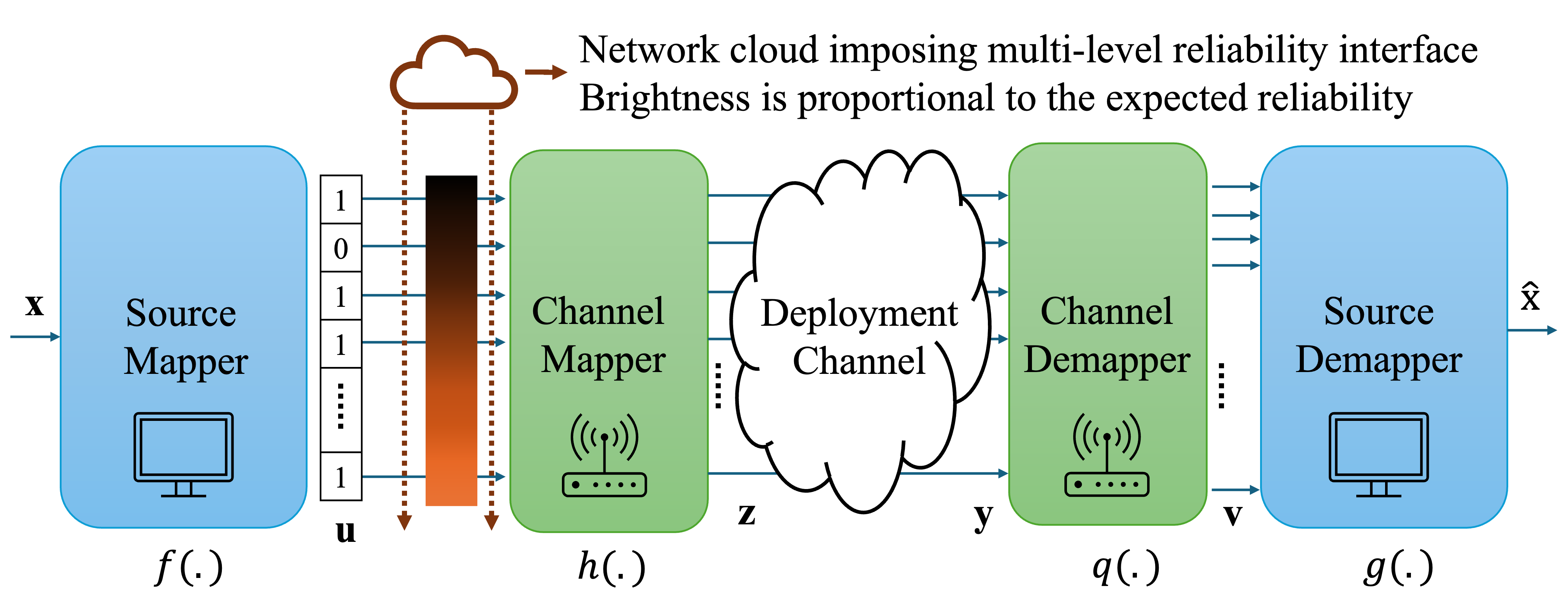}
    \caption{Diagram illustrating how application and network providers can interoperate by adhering to the proposed binary multilevel reliability interface, eliminating the need for direct communication or synchronization, and thus realizing the benefits of \gls{jscc} in practice.\vspace{-0.3cm}}
    \label{fig:interface_idea}
\end{figure}

In our model, both the application provider and network provider aim to control the design of their coding schemes independently, without requiring a synchronized training process. 
Thus, we consider the design of \gls{dl}-driven source and channel codecs separated via a novel binary interface. This binary interface serves to decouple the design of source and channel codecs using a shared guideline, rather than direct communication or joint design, Fig.~\ref{fig:interface_idea}. We term the mapping from the source to the binary codeword as ``source mapping" and the mapping from the binary codeword to channel input symbols as ``channel mapping". 
The source mapper learns to partition the content based on the reliability of bit locations, assigning less information to unreliable bit locations and more to reliable ones. On the other hand, the channel mapper learns to adjust the channel distribution to match that for which the source mapper is trained. 

In our setup, to conform to the binary interface, the source mapper is trained independently for a channel comprising multiple noisy bit pipes, each with a distinct bit-flipping probability, referred to as the ``multi-level \gls{ber} medium" henceforth. This modeling acts as an implicit, logical medium for conveying the relative importance of each bit to the channel mapper.
This effectively trains the source mapper to function as a binary \gls{jscc} without real-time information about the wireless channel.
Our proposed approach can be readily implemented in communication networks by packetizing the source bits corresponding to each reliability level, with the header indicating the level index.
The \gls{ber} values corresponding to the different levels can be standardized or communicated by the source to the network before the start of the session. After training the source mapper and demapper, we freeze them and proceed to train the channel mapper and demapper. This training utilizes the distortion observed from the source demapper over a wireless channel to map the source codeword to channel input symbols and vice versa. To illustrate the concept, we focus on \gls{awgn} channels in this paper. By updating the weights of the channel mapper and demapper using the application distortion measure, the channel mapper learns to align the \gls{awgn} channel distribution with the multi-level reliability medium for which the source mapper is trained.



The current setup bears similarity with two types of existing work: communication over relay channels and \gls{uep}. Existing research on semantic communication over relay channels has been investigated over multiple wireless links, e.g., \cite{bi17_joint_sourc_chann_codin_jpeg, bian22_deep_joint_sourc_chann_codin}, to leverage spatial diversity for enhanced resource allocation. However, in these approaches, the mapping from source bits to channel symbols still occurs at the source, raising practical concerns. \gls{uep} has been employed as part of \gls{jscc} for video transmission, as seen in \cite{kondi_joint_2001, ji_joint_2012}. However, these methods are typically tailored to specific video codecs, and their performance is constrained by the protocol's flexibility. Some recent work also considers the problem of digital semantic communication, as evidenced by \cite{fu23_vector_quant_seman_commun_system}; however, none fully satisfy the requirements outlined in this paper regarding applicability to current communication networks.


To summarize, the primary contributions of this paper are:
\begin{enumerate}
        \item We introduce a multi-level \gls{ber} interface as an abstraction of the underlying channel, enabling the source to learn a binary \gls{jscc} that adheres to a wide range of channel conditions. The proposed multi-level \gls{ber} interface also acts as an abstraction of the source indicating multiple levels of importance. 
        \item As a beneficiary of the proposed scheme, we devise a two-step training process: The source mapper and demapper undergo training over the multi-level \gls{ber} medium. Then, the channel mapper and demapper are trained by querying the source demapper. 
        The proposed scheme, which we call \emph{Split DeepJSCC}, achieves lower end-to-end distortion and exhibits graceful degradation of distortion with varying channel quality. 
\end{enumerate}
Our proposed scheme differs from source coding with successive refinement or multi-level source coding found in the literature (e.g., \cite{Cover1991, Rose2003, SuccessiveRefinementImageDeep, Harish2000, Foad2022}) as it involves joint source-channel coding at two distinct locations in the network, without the need for a joint design. One location directly accesses the source data, while the other accesses the channel.



\section{Problem formulation}
\label{sec:problem_def}

We consider the problem of wireless image transmission over an \gls{awgn} channel,\footnote{Note that this framework is applicable to more general wireless channels, including fading channels and channels with memory or feedback.} where the source and channel are separated by a binary interface, as depicted in Fig.~\ref{fig:interface_idea}. Both the application provider and network provider aim to control the design of their coding schemes independently, without requiring a synchronized training process.

Let the image $\mathbf{x} \in \{ 0, ..., 255 \}^{H \times W \times C}$ be encoded by a source mapper $f: \{ 0, ..., 255 \}^{H \times W \times C} \mapsto \{ 0, 1 \}^M$ into a binary format, where $H$, $W$, and $C$ represent the image's height, width, and color channels, respectively ($C=3$ for RGB images). A channel mapper $h: \{ 0, 1 \}^M \mapsto \mathbb{C}^K$ then maps the output of the source mapper, denoted as $\mathbf{u} = f(\mathbf{x})$, to $K$ complex channel input symbols $\mathbf{z} = h(\mathbf{u})$. We impose an average transmit power constraint $\bar{P}$, such that
\begin{equation}
\frac{1}{K} \sum_{i = 1}^K |z_i|^2 \leq \bar{P},
\end{equation}
where $z_i$ is the $i$-th element of the vector $\mathbf{z}$. The channel input symbols $\mathbf{z}$ are then transmitted through an \gls{awgn} channel $W: \mathbb{C}^K \mapsto \mathbb{C}^K$, such that $\mathbf{y} = W(\mathbf{z}) = \mathbf{z} + \mathbf{n}$, where $\mathbf{n} \sim CN(0, \sigma^2 \mathbf{I}_{K \times K})$ is an \gls{iid} complex Gaussian vector.

At the receiver, a channel demapper $q: \mathbb{C}^K \mapsto \{ 0, 1 \}^M$ maps the channel output $\mathbf{y}$ to a binary vector $\mathbf{v}$, so that $\mathbf{v} = q( \mathbf{y} )$.
Finally, a source demapper $g: \{0, 1\}^M \mapsto \{ 0, ..., 255 \}^{H \times W \times C}$ maps the binary vector $\mathbf{v}$ to an estimate of the source image $\hat{ \mathbf{x} } = g( \mathbf{v} )$.
The key difference between the proposed setup and the traditional separate source and channel coding problem lies in our focus on achieving the best possible reconstruction $\hat{\mathbf{x}}$ for a given number of channel uses $K$, rather than optimizing the accuracy of the received bits $\mathbf{v}$ and the reconstruction $\hat{\mathbf{x}}$ separately.
In the above problem formulation, we define the channel \gls{snr} as follows,
\begin{equation}
    \text{SNR} = 10\log_{10} \left( \frac{\bar{P}}{{\sigma}^2} \right) \text{ dB}.
\end{equation}
We further define the \emph{\gls{bcr}}, denoted by $\rho$, as
\begin{equation}
    \rho = \frac{K}{H \times W \times C}\cdot
\end{equation}

The quality of the received image is measured by \gls{psnr},
\begin{equation}
    \text{PSNR}(\mathbf{x},\hat{\mathbf{x}})=10\log_{10}\bigg(\frac{A^2}{\text{MSE}(\mathbf{x},\hat{\mathbf{x}})}\bigg)~\text{dB},
    \label{eq:psnr_def}
\end{equation}
where $A$ is the maximum value taken by each element of $\mathbf{x}$ ($255$ for an 8-bit RGB pixel), and $ \text{MSE}(\mathbf{x},\hat{\mathbf{x}})=||\mathbf{x}-\hat{\mathbf{x}}||_2^2$ is the \gls{mse} between the original image $\mathbf{x}$ and the reconstruction $\hat{\mathbf{x}}$.

\section{Proposed solution}
\label{sec:proposed_soln}

In this paper, we address the problem outlined in Sec. \ref{sec:problem_def} via parameterizing the source mapper and demapper and channel mapper and demapper, using \glspl{dnn}.
Specifically, the image $\mathbf{x}$ is mapped to a binary representation $\mathbf{u} = f_{\boldsymbol{\theta}}( \mathbf{x} )$ using a non-linear encoder $f_{\boldsymbol{\theta}}: \{0, ..., 255\}^{H \times W \times C} \mapsto \{0, 1\}^M$, where $\boldsymbol{\theta}$ denotes the \gls{dnn} weights.
Similarly, the channel mapper $h_{\boldsymbol{\psi}}: \{0, 1\}^M \mapsto \mathbb{C}^K$ maps $\mathbf{u}$ to the channel input $\mathbf{z}$, the channel demapper $q_{\boldsymbol{\phi}}: \mathbb{C}^K \mapsto \{0, 1\}^M$ maps the channel output $\mathbf{y}$ to the binary vector $\mathbf{v}$, and the source demapper $g_{\boldsymbol{\eta}}: \{0, 1\}^M \mapsto \{ 0, ..., 255 \}^{H \times W \times C}$ maps $\mathbf{v}$ back to the source domain $\hat{\mathbf{x}}$.

The primary challenge in using \gls{dnn} to parameterize each of the aforementioned functions is the non-differentiability of mapping real-valued vectors to binary vectors. To address this issue, we propose a two-step training process, employing variational learning techniques in each step to cast the binary mapping as a generative process. To this end, during the training process, the source mapper (resp., channel demapper) is considered a probabilistic encoder $p( \mathbf{u}^\prime | \mathbf{x} ,\boldsymbol{\theta})$ (resp., $p( \mathbf{v}^\prime | \mathbf{y},\boldsymbol{\phi} )$), with outputs drawn from independent Bernoulli distributions, where the function $f_{\boldsymbol{\theta}}( \cdot )$ (resp., $q_{\boldsymbol{\phi}}(\mathbf{y})$) indicates the probability of the corresponding bit being $1$. Once trained, these probabilistic encoders are replaced with their deterministic maximum likelihood versions to become binary encoders via an element-wise rounding operation, i.e.,
\begin{equation}\label{eq:determntic_encoders}
    \begin{split}
        {\mathbf{u}} &= \argmax_{\mathbf{u}^\prime} p( \mathbf{u}^\prime | \mathbf{x},\boldsymbol{\theta} ) =  \lfloor f_{\boldsymbol{\theta}}(\mathbf{x}) \rceil,\\
        {\mathbf{v}} &= \argmax_{\mathbf{v}^\prime} p( \mathbf{v}^\prime | \mathbf{y},\boldsymbol{\phi} ) = \lfloor q_{\boldsymbol{\phi}}(\mathbf{y}) \rceil.     
    \end{split}
\end{equation}

\subsection{Training Source Mapper and Demapper}
\label{subsec:source_pretraining}

\begin{figure}
    \centering
    \includegraphics[width=0.8\linewidth]{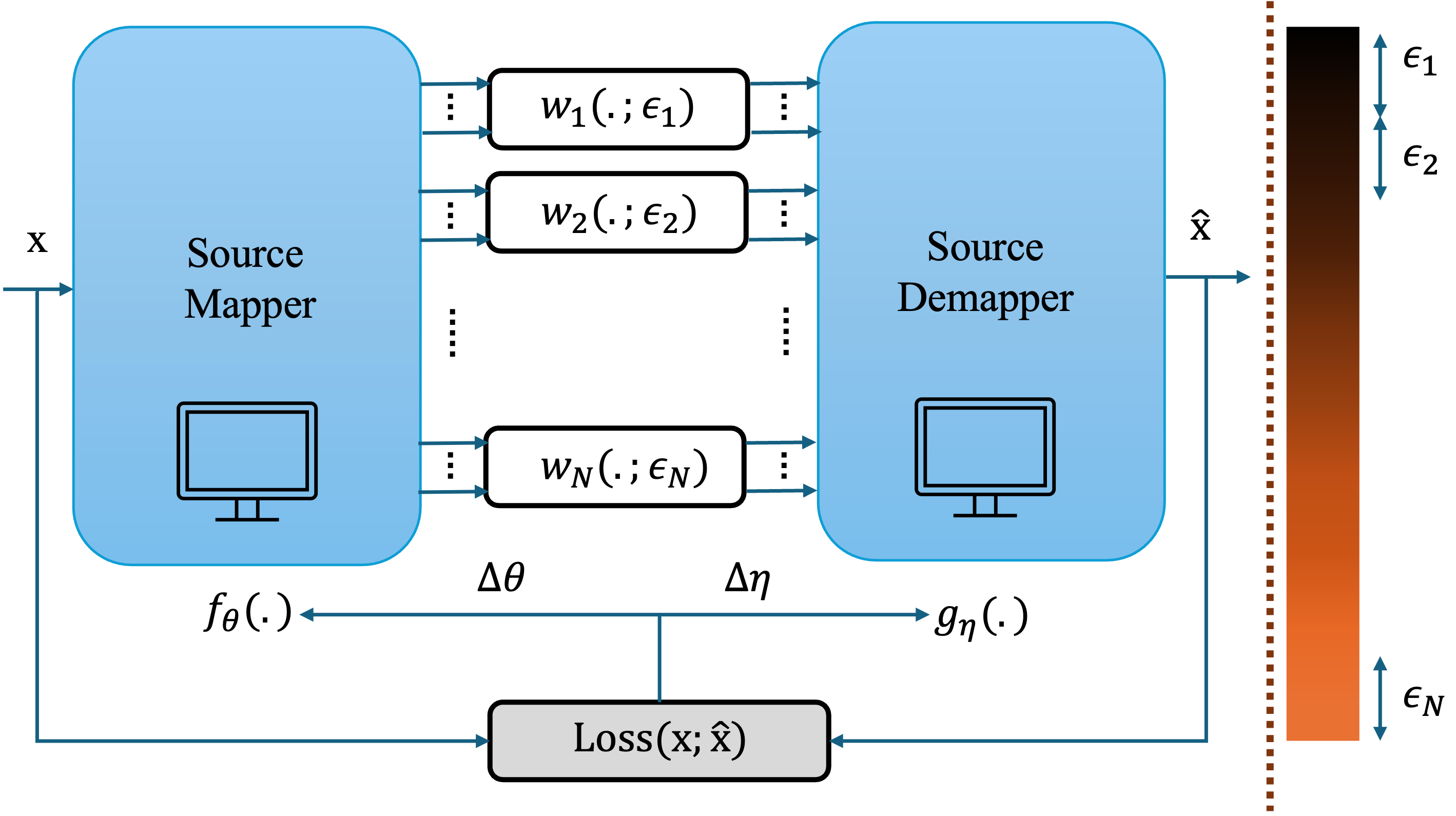}
    \caption{Diagram depicting training process of the source mapper and demapper. The blue boxes are trainable components of the architecture.}
    \label{fig:source_pretraining}
\end{figure}

We first train the source mapper $f_{\boldsymbol{\theta}}$ and demapper $g_{\boldsymbol{\eta}}$ with a multi-level \gls{ber} medium in between. 
Utilizing the trained source codec, images are mapped to binary representations, and noisy binary representations are mapped to reconstructed images.
We ensure that the binary representation conforms to the multi-level \gls{ber} interface by modeling the medium as a series of $N<<M$ parallel \glspl{bsc} with different error probabilities, conveying the expected reliability level for each group of $M/N$ bits to the source mapper.
In other words, let $\{ {w}_i( \cdot , \epsilon_i ) \}_{i=1}^N$ represent $N$ independent \glspl{bsc} with error probability $\epsilon_i$, defined as 
\begin{equation}
    {w}_i( u , \epsilon_i) = \begin{cases}
        u \oplus 1, ~ &\text{w.p. } \epsilon_i, \\
        u, ~ &\text{w.p. } 1 - \epsilon_i,
    \end{cases}
\end{equation}
where $\oplus$ denotes the binary (modulo 2) addition operation.\footnote{We intentionally use different notation for the deployment channel $W(\cdot)$ and the medium $\bar{\mathbf{w}}( \cdot, \boldsymbol{\epsilon})$ used for training the source codec.} A diagram illustrating the training process for the source codec is shown in Fig.~\ref{fig:source_pretraining}.

We now formulate the representation learning of $\mathbf{u}$. Without loss of generality and for simplicity, we assume $M$ is divisible by $N$. Let $\bar{\mathbf{w}}( \cdot, \boldsymbol{\epsilon}) = [w_1( \cdot, \epsilon_1), \ldots, w_N( \cdot, \epsilon_N)]$ and $\hat{\mathbf{u}} = \bar{\mathbf{w}}( \mathbf{u}, \boldsymbol{\epsilon})$ be the output of the multi-level \glspl{ber} medium. Here, the first $M/N$ bits pass through $w_1$, the second $M/N$ bits pass through $w_2$, and so forth. The \gls{jscc} objective for training the source mapper is to identify $\boldsymbol{\theta}^*$, such that
\begin{equation*}
    \boldsymbol{\theta}^*=\argmax_{\boldsymbol{\theta}} I( \mathbf{x}; \hat{\mathbf{u}}| \boldsymbol{\theta}).
\end{equation*}
The function $I(\cdot ;\cdot)$ outputs the mutual information between two random variables. 

By the definition of the mutual information and since $\mathbf{x}$ does not depend on $\boldsymbol{\theta}$, we have
\begin{equation*}
    \begin{split}
        \boldsymbol{\theta}^*&=\argmax_{\boldsymbol{\theta}} H(\mathbf{x}) - H(\mathbf{x} | \hat{\mathbf{u}}, \boldsymbol{\theta})=\argmax_{\boldsymbol{\theta}}- H(\mathbf{x} | \hat{\mathbf{u}}, \boldsymbol{\theta})\\
        &=\argmax_{\boldsymbol{\theta}} E_{\mathbf{x}}E_{\hat{\mathbf{u}} \sim p(\hat{\mathbf{u}} | \mathbf{x}, \boldsymbol{\theta})} \left[ \log {p}( \mathbf{x} | \hat{\mathbf{u}}, \boldsymbol{\theta}) \right],
    \end{split}
\end{equation*}
where ${p}( \mathbf{x} | \hat{\mathbf{u}}, \boldsymbol{\theta})$ is the true posterior -- representing the posterior probability over all possible images $\mathbf{x}$ given the received noisy codeword $\hat{\mathbf{u}}$ and the encoder's parameter $\boldsymbol{\theta}$. This posterior represents the best possible decoder. Since the posterior distribution is intractable to evaluate and use for optimization, we instead use a variational posterior approximation parameterized by $\boldsymbol{\eta}$ as $\tilde{p}( \mathbf{x} | \hat{\mathbf{u}},\boldsymbol{\theta},\boldsymbol{\eta})$. This approximation acts as a probabilistic decoder for $\hat{\mathbf{u}}$, as proposed in, e.g., \cite{KingmaWellingICLR2014,rezende2014stochastic}. Thus, maximizing the tractable likelihood leads to an approximation of the best decoder. Consequently, the joint process of training the source mapper and demapper is formulated as,
\begin{equation*}
        \boldsymbol{\theta}^*,\boldsymbol{\eta}^*=\argmax_{\boldsymbol{\theta},\boldsymbol{\eta}}  E_{\mathbf{x}}E_{\hat{\mathbf{u}} \sim p(\hat{\mathbf{u}} | \mathbf{x}, \boldsymbol{\theta})} \left[  \log \tilde{p}( \mathbf{x} | \hat{\mathbf{u}}, \boldsymbol{\theta},\boldsymbol{\eta}) \right].
\end{equation*}
Hence, the objective for the training process is,
\begin{equation*}
    \mathcal{L}=E_{\mathbf{x}}E_{\hat{\mathbf{u}} \sim p(\hat{\mathbf{u}} | \mathbf{x}, \boldsymbol{\theta})} \left[  \log \tilde{p}( \mathbf{x} | \hat{\mathbf{u}}, \boldsymbol{\theta},\boldsymbol{\eta}) \right].
\end{equation*}

To mitigate the variance of the stochastic approximation $\tilde{p}( \mathbf{x} | \hat{\mathbf{u}}, \boldsymbol{\theta},\boldsymbol{\eta})$, we adopt the multi-sample technique proposed in, e.g., \cite{burda2016importance,bornschein2015reweighted}, wherein we replace the approximation with its average from $J>1$ instances. This modification results in the following adjusted loss function,
\begin{equation*}
    \mathcal{L}\approx E_{\mathbf{x}}E_{\hat{\mathbf{u}}^{1:J} \sim p(\hat{\mathbf{u}}^{1:J} | \mathbf{x}, \boldsymbol{\theta})} \left[  \log \frac{1}{J}\sum_{j=1}^J  \tilde{p}( \mathbf{x} | \hat{\mathbf{u}}^j, \boldsymbol{\theta},\boldsymbol{\eta}) \right].
\end{equation*}
Here, $\hat{\mathbf{u}}^{1:J}$ represent $J$ samples independently drawn from $p(\hat{\mathbf{u}} | \mathbf{x}, \boldsymbol{\theta})$, thus ${p(\hat{\mathbf{u}}^{1:J} | \mathbf{x}, \boldsymbol{\theta})}=\prod_{j=1}^J p(\hat{\mathbf{u}}^j | \mathbf{x}, \boldsymbol{\theta})$. We then replace the first expectation with an empirical average over a source dataset $\mathcal{D}$ to obtain a Monte Carlo objective, as $\mathcal{L}^J=\frac{1}{|\mathcal{D}|}\sum_{\mathbf{x}\in\mathcal{D}} \mathcal{L}^J(\mathbf{x})$, where
\begin{equation}\label{eq:cost-per-sample}
    \mathcal{L}^J(\mathbf{x})\triangleq E_{\hat{\mathbf{u}}^{1:J} \sim p(\hat{\mathbf{u}}^{1:J} | \mathbf{x}, \boldsymbol{\theta})} \left[\log \frac{1}{J}  \sum_{j=1}^J \tilde{p}( \mathbf{x} | \hat{\mathbf{u}}^j, \boldsymbol{\theta},\boldsymbol{\eta}) \right].
\end{equation}

Next, we focus on $\mathcal{L}^J(\mathbf{x})$ and derive its gradients with respect to $\theta$ and $\eta$ to optimize the source code's parameters using the Stochastic Gradient Descent (SGD) method. It is demonstrated in Appendix~\ref{sec:appendix_grad} that,
\begin{align}     
\nabla_{\boldsymbol{\theta}}\mathcal{L}^J(\mathbf{x})
    &{=}E_{\hat{\mathbf{u}}^{1:J}}\left[\hat{L}( \hat{\mathbf{u}}^{1:J} )\nabla_{\boldsymbol{\theta}} \log p(\hat{\mathbf{u}}^{1:J} | \mathbf{x}, \boldsymbol{\theta})\right]\label{eq:grad_theta_orig}\\
    &{+}E_{\hat{\mathbf{u}}^{1:J}}\left[\sum_{j=1}^J c_j(\hat{\mathbf{u}}^{1:J})\nabla_{\boldsymbol{\theta}} \log \tilde{p}( \mathbf{x} | \hat{\mathbf{u}}^j, \boldsymbol{\theta},\boldsymbol{\eta})\right],\nonumber\\      \nabla_{\boldsymbol{\eta}}\mathcal{L}^J(\mathbf{x})&{=}E_{\hat{\mathbf{u}}^{1:J}}[\sum_{j=1}^J c_j(\hat{\mathbf{u}}^{1:J})\nabla_{\boldsymbol{\eta}} \log \tilde{p}( \mathbf{x} | \hat{\mathbf{u}}^j, \boldsymbol{\theta},\boldsymbol{\eta})],\label{eq:grad_eta_orig}
\end{align}
where $\hat{\mathbf{u}}^{1:J}\sim p(\hat{\mathbf{u}}^{1:J} | \mathbf{x}, \boldsymbol{\theta})$. The helper functions $c_j(\hat{\mathbf{u}}^{1:J})$ and $\hat{L}( \hat{\mathbf{u}}^{1:J} )$ are given below below,
\begin{equation}\label{eq:helper_functions}
    \begin{split}
    \hat{L}( \hat{\mathbf{u}}^{1:J} ) &\triangleq \log \frac{1}{J} \sum_{j=1}^J \tilde{p}( \mathbf{x} | \hat{\mathbf{u}}^j , \boldsymbol{\theta},\boldsymbol{\eta}),\\
    c_j(\hat{\mathbf{u}}^{1:J})& \triangleq \frac{\tilde{p}( \mathbf{x} | \hat{\mathbf{u}}^j , \boldsymbol{\theta},\boldsymbol{\eta})}{\sum_{i=1}^J \tilde{p}( \mathbf{x} | \hat{\mathbf{u}}^i , \boldsymbol{\theta},\boldsymbol{\eta})}\cdot
    \end{split}
\end{equation}

Estimating $\nabla_{\boldsymbol{\theta}}\mathcal{L}^J(\mathbf{x})$ in (\ref{eq:grad_theta_orig}) is known to be challenging in practice, particularly when the latent space is discrete. This difficulty arises mainly due to the coefficient $\hat{L}( \hat{\mathbf{u}}^{1:J} )$ appearing in the first summation, because (i) the gradients for all $J$ samples are multiplied by the same coefficient, neglecting their individual importance, and (ii) the coefficient can potentially be unbounded, leading to unstable training. In \cite{mnih16_variat_monte_carlo}, it is proposed to replace $\hat{L}( \hat{\mathbf{u}}^{1:J} )$ with $\hat{L}( \hat{\mathbf{u}}^{j}| \hat{\mathbf{u}}^{-j})$ in a variational autoencoder structure to address these issues. The resulting objective is referred to as the \gls{vimco} estimator, and it will be employed in our paper,
\begin{equation*}
    \begin{split}        \nabla_{\boldsymbol{\theta}}\mathcal{L}^J(\mathbf{x})&=E_{\hat{\mathbf{u}}^{1:J}}\left[\sum_{j=1}^J \hat{L}( \hat{\mathbf{u}}^j | \hat{\mathbf{u}}^{-j} )\nabla_{\boldsymbol{\theta}} \log p(\hat{\mathbf{u}}^{j} | \mathbf{x}, \boldsymbol{\theta})\right]\\
    &+E_{\hat{\mathbf{u}}^{1:J}}\left[\sum_{j=1}^J c_j(\hat{\mathbf{u}}^{1:J})\nabla_{\boldsymbol{\theta}} \log \tilde{p}( \mathbf{x} | \hat{\mathbf{u}}^j, \boldsymbol{\theta},\boldsymbol{\eta})\right].
    \end{split}
\end{equation*}

The adjustment $\hat{L}( \hat{\mathbf{u}}^j | \hat{\mathbf{u}}^{-j} )$ is a function of $\hat{\mathbf{u}}^{-j}$, denoting all samples in $\hat{\mathbf{u}}^{1:J}$ except for $\hat{\mathbf{u}}^{j}$, and it is defined as follows,
\begin{equation*}
    \begin{split}
    \hat{L}( \hat{\mathbf{u}}^j | \hat{\mathbf{u}}^{-j} )&\triangleq \hat{L}( \hat{\mathbf{u}}^{1:J} ) \\
    &- \log \frac{1}{J} \sum_{i \ne j} \tilde{p}( \mathbf{x} | \hat{\mathbf{u}}^i, \boldsymbol{\theta},\boldsymbol{\eta}) + m( \mathbf{x} | \hat{\mathbf{u}}^{-j} , \boldsymbol{\eta} ),
    \end{split}
\end{equation*}
where $m( \mathbf{x} | \hat{\mathbf{u}}^{-j} , \boldsymbol{\eta} ) = \exp \left( \frac{1}{J - 1} \sum_{i \ne j} \log \tilde{p}( \mathbf{x} | \hat{\mathbf{u}}^i ,\boldsymbol{\theta}, \boldsymbol{\eta}) \right)$.

Next, the expectations are replaced with their one-sample realizations,
\begin{equation}\label{eq:grad_theta_eta}
    \begin{split}        \nabla_{\boldsymbol{\theta}}\mathcal{L}^J(\mathbf{x})&\approx\sum_{j=1}^J \hat{L}( \hat{\mathbf{u}}^j | \hat{\mathbf{u}}^{-j} )\nabla_{\boldsymbol{\theta}} \log p(\hat{\mathbf{u}}^{j} | \mathbf{x}, \boldsymbol{\theta})\\
    &+\sum_{j=1}^J c_j(\hat{\mathbf{u}}^{1:J})\nabla_{\boldsymbol{\theta}} \log \tilde{p}( \mathbf{x} | \hat{\mathbf{u}}^j, \boldsymbol{\theta},\boldsymbol{\eta}),\\       \nabla_{\boldsymbol{\eta}}\mathcal{L}^J(\mathbf{x})&\approx\sum_{j=1}^J c_j(\hat{\mathbf{u}}^{1:J})\nabla_{\boldsymbol{\eta}} \log \tilde{p}( \mathbf{x} | \hat{\mathbf{u}}^j, \boldsymbol{\theta},\boldsymbol{\eta}),
    \end{split}
\end{equation}
where $\hat{\mathbf{u}}^{1:J}\sim p(\hat{\mathbf{u}}^{1:J} | \mathbf{x}, \boldsymbol{\theta})$. 

The final step involves modeling $\log p(\hat{\mathbf{u}}^{j} | \mathbf{x}, \boldsymbol{\theta})$ and $ \log \tilde{p}( \mathbf{x} | \hat{\mathbf{u}}^j, \boldsymbol{\theta},\boldsymbol{\eta})$. We remind that the probabilistic encoder $p(\hat{\mathbf{u}} | \mathbf{x}, \boldsymbol{\theta})$ is related to the source mapper $f_{\boldsymbol{\theta}}$, where  $f_{\boldsymbol{\theta}}( \mathbf{x} )$ produces the mean of independent Bernoulli random variables representing the bits of $\mathbf{u}$, and we employ the sigmoid activation function to ensure that $f_{\boldsymbol{\theta}}( \mathbf{x} )$ falls within the range $[0, 1]^M$. The distribution $\tilde{p}( \mathbf{x} | \hat{\mathbf{u}}^j, \boldsymbol{\theta},\boldsymbol{\eta})$ is related to the source demapper $g_{\boldsymbol{\eta}}$, where $\hat{\mathbf{u}}$ represents the noisy version of $\mathbf{u}$.

\begin{lemma} \label{lemma:encoder_decoder}
The log likelihood $\log p( \hat{\mathbf{u}}|\mathbf{x},\boldsymbol{\theta})$ can be formulated as,
\begin{equation*}
    \begin{split}
        &\log p( \hat{\mathbf{u}}|\mathbf{x},\boldsymbol{\theta}) \\
    &=\sum_{i = 1}^N \sum_{j = (i - 1)M/N + 1}^{iM/N} \hspace{-0.3cm}{\hat{\mathbf{u}}_j}\log (f_{\boldsymbol{\theta}}( \mathbf{x} )_j(1-2\epsilon_i)+\epsilon_i)\\
    &+\sum_{i = 1}^N \sum_{j = (i - 1)M/N + 1}^{iM/N}\hspace{-0.3cm}({1-\hat{\mathbf{u}}_j})\log((1-\epsilon_i)+f_{\boldsymbol{\theta}}( \mathbf{x} )_j(2\epsilon_i-1)),\\
    &\log \tilde{p}( \mathbf{x} | \hat{\mathbf{u}}, \boldsymbol{\theta},\boldsymbol{\eta})\approx -|| \mathbf{x} - g_{\boldsymbol{\eta}}( \hat{\mathbf{u}} ) ||_2^2.
    \end{split}
\end{equation*}
\end{lemma}
The proof appears in the Appendix~\ref{sec:enc_dec_details}.

Using Eqns.~(\ref{eq:grad_theta_eta}) and Lemma~\ref{lemma:encoder_decoder},
each epoch of training the source mapper $f_{\boldsymbol{\theta}}$ and demapper $g_{\boldsymbol{\eta}}$ can be described as follows,
\begin{equation}
\begin{split}
\boldsymbol{\theta}&:=\boldsymbol{\theta}+\frac{1}{|\mathcal{D}|}\sum_{\mathbf{x}\in\mathcal{D}}\nabla_{\boldsymbol{\theta}}\mathcal{L}^J(\mathbf{x}),\\
\boldsymbol{\eta}&:=\boldsymbol{\eta}+\frac{1}{|\mathcal{D}|}\sum_{\mathbf{x}\in\mathcal{D}}\nabla_{\boldsymbol{\eta}}\mathcal{L}^J(\mathbf{x}).
\end{split}
\end{equation}

\subsection{Training Channel Mapper and Demapper}
\label{subsec:e2e_training}

\begin{figure}
    \centering
    \includegraphics[width=\linewidth]{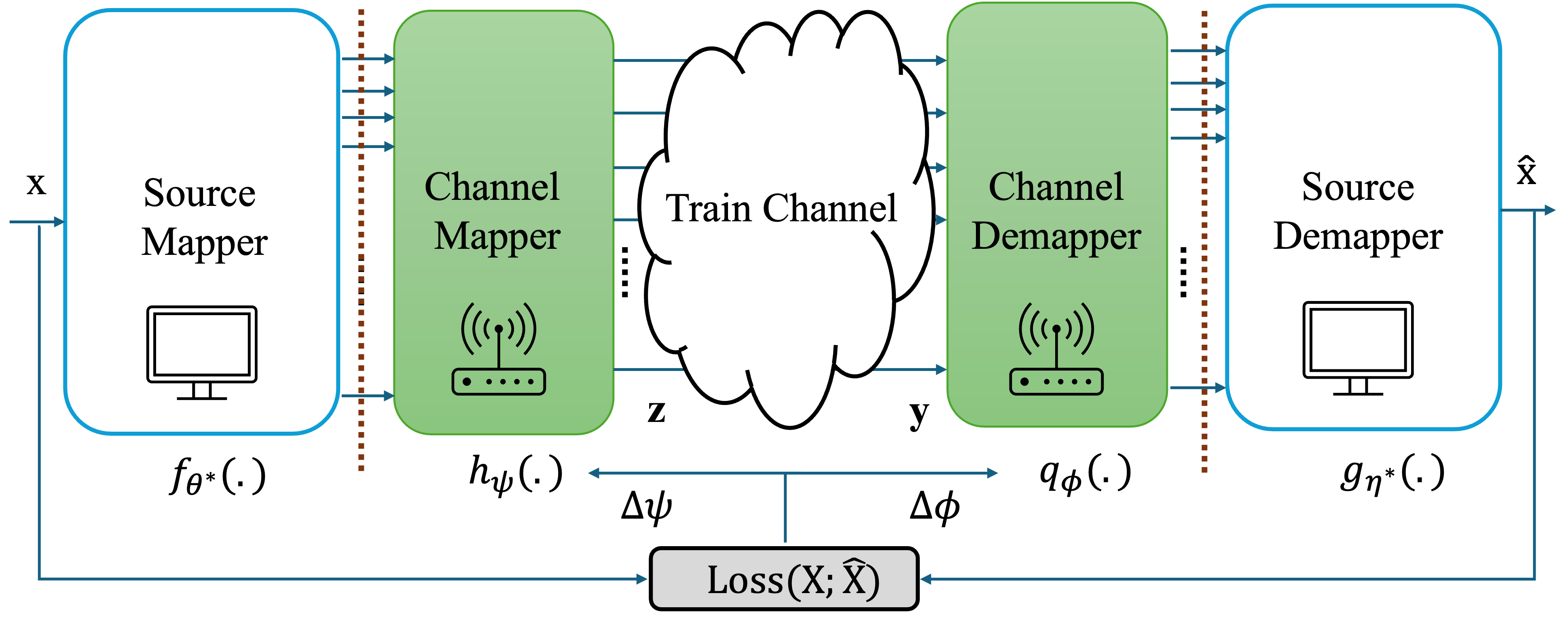}
    \caption{Diagram illustrating the training process of channel mapper and demapper. The green boxes are trainable parts of the architecture.}
    \label{fig:e2e_training}
\end{figure}

After training the source mapper $f_{\boldsymbol{\theta}}$ and demapper $g_{\boldsymbol{\eta}}$, we freeze their weights and replace the the probabilistic encoder $p( \mathbf{u} | \mathbf{x} )$ with the deterministic \gls{ml} estimator as described in (\ref{eq:determntic_encoders}). We then proceed to train the channel mapper and demapper as another variational learning problem, with an \gls{awgn} channel in between. 
A diagram illustrating this setup is shown in Fig.~\ref{fig:e2e_training}. 

The channel mapper maps the codeword ${\mathbf{u}}=\lfloor f_{\boldsymbol{\theta}}(\mathbf{x}) \rceil$ to $2K$ channel input symbols in $\mathbf{z} = h_{\boldsymbol{\psi}}( \mathbf{u} )$, and then consecutive values are paired to form $K$ complex-valued symbols by mapping elements of each pair to the real and imaginary parts of a complex symbol.
To ensure that the channel input symbols meet the average power constraint, we perform normalization as
\begin{equation}
    z_i \leftarrow \sqrt{ \frac{K \bar{P} }{ \mathbf{z}^H \mathbf{z} } } z_i,
\end{equation}
where $z_i$ is the $i$-th value in $\mathbf{z}$ and $\mathbf{z}^H$ is the Hermitian transpose of $\mathbf{z}$.

The normalized channel input symbols $\mathbf{z}$ are then transmitted across an \gls{awgn} channel $W_\text{Train}$, and the received value $\mathbf{y} = W_\text{Train}( \mathbf{z} )$ is passed to the channel demapper $q_{\boldsymbol{\phi}}$.
We model the channel demapper using the same stochastic formulation as for training the source mapper to achieve a binary representation for the output while mitigating the non-differentiability issue.
That is, we model the conditional distribution (probabilistic decoder) $p( \mathbf{v} | \mathbf{y} )$ as
\begin{equation}
    p( \mathbf{v} | \mathbf{y} ) = \prod_{j=1}^M q_{\boldsymbol{\phi}}( \mathbf{y} )_j^{v_j} (1 - q_{\boldsymbol{\phi}}( \mathbf{y} )_j )^{1 - v_j},
\end{equation}
where $g_{\boldsymbol{\phi}}( \mathbf{y} )_j$ denotes the $j$-th output of $q_{\boldsymbol{\phi}}( \mathbf{y} )$, and we use the sigmoid activation function to ensure that $q_{\boldsymbol{\phi}}( \mathbf{y} )$ falls within the range $[0, 1]^M$.

We note that $q_{\boldsymbol{\phi}}( \mathbf{y} )$ resembles a traditional channel decoder, where the output of the decoding algorithm is typically a soft decision (a probability) that each bit is either a $0$ or $1$.
However, in contrast to a traditional channel decoder, our priority is not exact reconstruction of $\mathbf{u}$.
Instead, we train $h_{\boldsymbol{\psi}}( \mathbf{u}^{\prime} )$ and $q_{\boldsymbol{\phi}}( \mathbf{y} )$ using the distortion observed at the image demapper $g_{\boldsymbol{\eta}}$ so that the channel mapper and demapper learn to encode the image codeword $\mathbf{u}$ based on the application objective.
This allows the channel mapper and demapper to be semantically aware while being decoupled from the source mapper and demapper.

We highlight that unlike the multi-level \gls{bsc} medium, the \gls{awgn} channel is differentiable.
Therefore, we directly learn the representation $\mathbf{v}$  at the output of $q_{\boldsymbol{\phi}}( \mathbf{y} )$, by treating the composition $q_{\boldsymbol{\phi}} \circ W_{\sigma} \circ h_{\boldsymbol{\psi}}$ as a single generative model. 
More specifically, the training objective is $\mathcal{L}^J=\frac{1}{|\mathcal{D}|}\sum_{\mathbf{x}\in\mathcal{D}}\mathcal{L}^J(\mathbf{x})$ for a source dataset $\mathcal{D}$, where
\begin{equation*}
    \begin{split}
    \mathcal{L}^J(\mathbf{x}){\triangleq} E_{{\mathbf{v}}^{1:J} \sim p({\mathbf{v}}^{1:J} | \lfloor f_{\boldsymbol{\theta}}(\mathbf{x}) \rceil, \boldsymbol{\psi},\boldsymbol{\phi})} \left[\log \frac{1}{J}  \sum_{j=1}^J \tilde{p}( \mathbf{x} | {\mathbf{v}}^j, \boldsymbol{\theta},\boldsymbol{\eta}) \right].
    \end{split}
\end{equation*}
We use the \gls{vimco} \cite{mnih16_variat_monte_carlo} method again to estimate the gradient of the loss function with respect to $\boldsymbol{\psi}$ and $\boldsymbol{\phi}$. 
\begin{equation}\label{eq:grad_channel_codec}
    \begin{split}
    &\nabla_{\boldsymbol{\psi}, \boldsymbol{\phi}}\mathcal{L}^J(\mathbf{x})\\
    &=\nabla_{\boldsymbol{\psi}, \boldsymbol{\phi}}E_{{\mathbf{v}}^{1:J} \sim p({\mathbf{v}}^{1:J} | \lfloor f_{\boldsymbol{\theta}}(\mathbf{x}) \rceil, \boldsymbol{\psi},\boldsymbol{\phi})} \left[\log \frac{1}{J}  \sum_{j=1}^J \tilde{p}( \mathbf{x} | {\mathbf{v}}^j, \boldsymbol{\theta},\boldsymbol{\eta}) \right] \\
    &=\sum_{{\mathbf{v}}^{1:J}}\nabla_{\boldsymbol{\psi}, \boldsymbol{\phi}}p({\mathbf{v}}^{1:J} | \lfloor f_{\boldsymbol{\theta}}(\mathbf{x}) \rceil, \boldsymbol{\psi},\boldsymbol{\phi}) [\log \frac{1}{J}  \sum_{j=1}^J \tilde{p}( \mathbf{x} | {\mathbf{v}}^j, \boldsymbol{\theta},\boldsymbol{\eta}) ] \\
    &= E_{{\mathbf{v}}^{1:J}} \left[\hat{L}( {\mathbf{v}}^{1:J})\nabla_{\boldsymbol{\psi}, \boldsymbol{\phi}} \log p(\mathbf{v}^{1:J} | \lfloor f_{\boldsymbol{\theta}}(\mathbf{x}) \rceil, \boldsymbol{\psi}, \boldsymbol{\phi}) )\right]\\
    &\approx \sum_{j=1}^J \hat{L}( \mathbf{v}^j | \mathbf{v}^{-j}) \nabla_{\boldsymbol{\psi}, \boldsymbol{\phi}} \log p(\mathbf{v}^j | \lfloor f_{\boldsymbol{\theta}}(\mathbf{x}) \rceil, \boldsymbol{\psi}, \boldsymbol{\phi}).
    \end{split}
\end{equation}
We have
\begin{equation*}
    \begin{split}
        &\hat{L}( {\mathbf{v}}^{1:J} ) {=} \log \frac{1}{J} \sum_{j=1}^J \tilde{p}( \mathbf{x} | {\mathbf{v}}^j,\boldsymbol{\theta},\boldsymbol{\eta}),\\
        &\hat{L}( {\mathbf{v}}^j | {\mathbf{v}}^{-j} ){=} \hat{L}({\mathbf{v}}^{1:J} )\\
        &\hspace{1.5cm}- \log \frac{1}{J} \sum_{i \ne j} \tilde{p}( \mathbf{x} | {\mathbf{v}}^i,\boldsymbol{\theta},\boldsymbol{\eta}) + m( \mathbf{x} | {\mathbf{v}}^{-j} , \boldsymbol{\theta},\boldsymbol{\eta} ),\\
        &m( \mathbf{x} | {\mathbf{v}}^{-j} , \boldsymbol{\eta} ) {=} \exp \left( \frac{1}{J - 1} \sum_{i \ne j} \log \tilde{p}( \mathbf{x} | {\mathbf{v}}^i,\boldsymbol{\theta}, \boldsymbol{\eta}) \right).
    \end{split}
\end{equation*}
We remind from Lemma~\ref{lemma:encoder_decoder} that 
\begin{equation*}
    \log \tilde{p}( \mathbf{x} | {\mathbf{v}},\boldsymbol{\theta}, \boldsymbol{\eta})\approx -|| \mathbf{x} - g_{\boldsymbol{\eta}}( {\mathbf{v}} ) ||_2^2.
\end{equation*}
Lemma~\ref{lemma:ch_encoder_decoder} quantifies the term $p(\mathbf{v} | \lfloor f_{\boldsymbol{\theta}}(\mathbf{x}) \rceil, \boldsymbol{\psi}, \boldsymbol{\phi})$.

\begin{lemma} \label{lemma:ch_encoder_decoder}
The likelihood $p(\mathbf{v} | \mathbf{u}, \boldsymbol{\psi}, \boldsymbol{\phi})$ can be formulated as fallows,
\begin{equation*}
\begin{split}
    &p(\mathbf{v} | \mathbf{u}, \boldsymbol{\psi}, \boldsymbol{\phi})=\\
    &E_{\mathbf{n}} \left[\prod_{j=1}^M q_{\boldsymbol{\phi}}(h_{\boldsymbol{\psi}}(\mathbf{u})+\mathbf{n}  )_j^{v_j} (1 - q_{\boldsymbol{\phi}}( h_{\boldsymbol{\psi}}(\mathbf{u})+\mathbf{n} )_j  )^{1 - v_j}\right],
\end{split}
\end{equation*}
where $\mathbf{n} \sim CN(0, \sigma_\text{Train}^2 \mathbf{I}_{K \times K})$ is an \gls{iid} complex Gaussian vector with dimension $K$.
\end{lemma}
The proof appears in the Appendix~\ref{sec:ch_enc_dec_details}.

This will complete the derivations to update the parameters of the channel mapper and demapper using the SGD method,
\begin{equation}\label{eq:channel_codec_updates}
\begin{split}
\boldsymbol{\psi}&:=\boldsymbol{\psi}+\frac{1}{|\mathcal{D}|}\sum_{\mathbf{x}\in\mathcal{D}}\nabla_{\boldsymbol{\psi}}\mathcal{L}^J(\mathbf{x}),\\
\boldsymbol{\phi}&:=\boldsymbol{\phi}+\frac{1}{|\mathcal{D}|}\sum_{\mathbf{x}\in\mathcal{D}}\nabla_{\boldsymbol{\phi}}\mathcal{L}^J(\mathbf{x}).
\end{split}
\end{equation}
We call the proposed design where the source code and channel code are trained independently using our introduced binary interface as \emph{Split DeepJSCC}.

\section{Experiments}
\label{sec:experiments}

In this section, we provide a comprehensive overview of the training methodology employed, the selected hyperparameters, the architectures of the \gls{dnn} models utilized, and the numerical outcomes obtained.

\begin{figure*}
    \centering
    \includegraphics[width=0.7\linewidth]{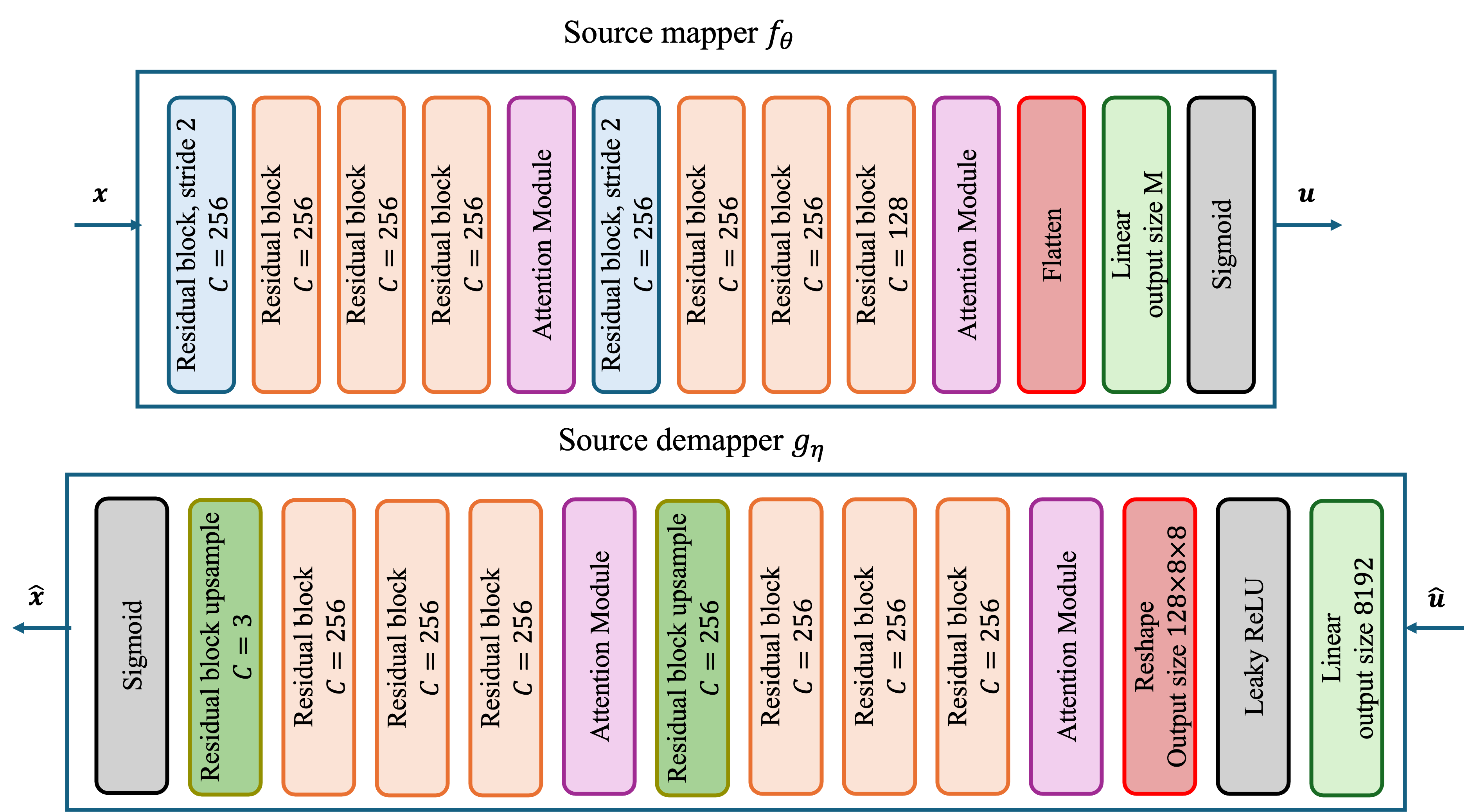}
    \caption{\gls{dnn} architectures used for source mapper ($f_{\boldsymbol{\theta}}$) and demapper ($g_{\boldsymbol{\eta}}$).}
    \label{fig:source_encoder_decoder_arch}
\end{figure*}

\begin{figure*}
    \centering
    \includegraphics[width=0.8\linewidth]{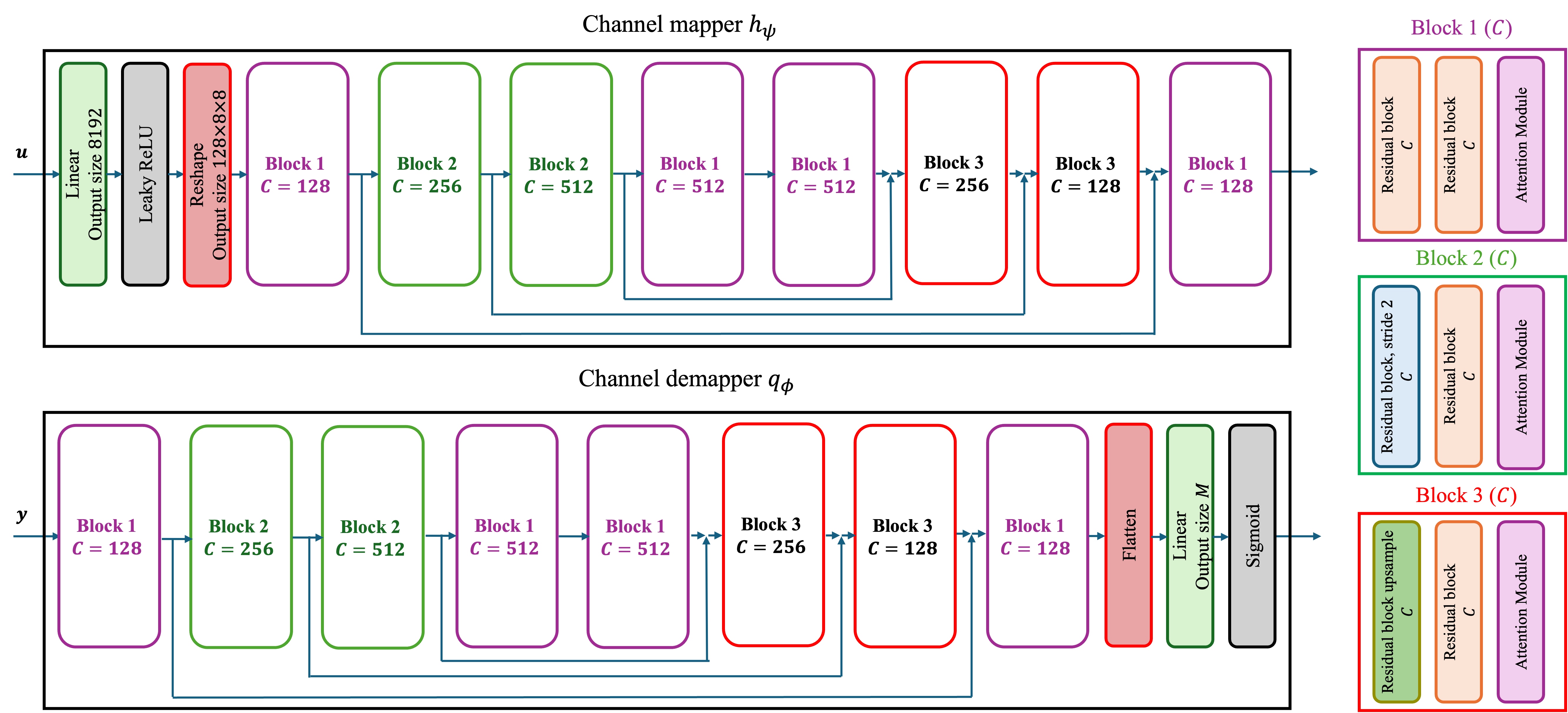}
    \caption{\gls{dnn} architectures for channel mapper ($h_{\boldsymbol{\psi}}$) and demapper ($q_{\boldsymbol{\phi}}$). The two arrows that merge into one represent the concatenation operation.}
    \label{fig:channel_mapper_demapper_arch}
\end{figure*}

\subsection{Training details}
\label{subsec:training_details}

We utilize the CIFAR10 dataset \cite{krizhevsky09_learn_multip_layer_featur_tiny_images}, which comprises $60{,}000$ images sized $32 \times 32 \times 3$.
The dataset is partitioned into training, validation, and evaluation sets in a $0.8:0.1:0.1$ ratio.
The \gls{dnn} architectures for the source mapper $f_{\boldsymbol{\theta}}$, demapper $g_{\boldsymbol{\eta}}$, channel mapper $h_{\boldsymbol{\psi}}$, and demapper $q_{\boldsymbol{\phi}}$ are depicted in Fig. \ref{fig:source_encoder_decoder_arch} and \ref{fig:channel_mapper_demapper_arch}, consisting of cascaded layers of residual blocks and attention blocks.
In these architectures, $C$ denotes the number of output channels for convolutional operations, such as those within a residual block.
The detailed descriptions of the constituent blocks can be found in CompressAI library \cite{begaint2020compressai} and the relevant publications, e.g., \cite{ballé15_densit_model_images_gener_normal_trans,cheng20_learn_image_compr_with_discr,shi16_real_time_singl_image_video}.

\begin{figure}
\pgfplotstableread[col sep=comma,]{bsc_pe.csv}\datatable
\begin{tikzpicture}
\centering
    \begin{axis}[
        ybar,
        bar width=.2cm,
        width=0.7\linewidth,
        height=0.4\linewidth,
        ymin=0.001,
        xtick=data,
        xticklabels from table={\datatable}{subchannel},
        ylabel={$\epsilon_i$},
        xlabel={subchannel index}]
        \addplot table [x expr=\coordindex, y={epsilon}]{\datatable};
    \end{axis}
\end{tikzpicture}
    \caption{The parameters of the multi-level \gls{ber} medium used for training the source codec.}
    \label{fig:subchannel_pe}
\end{figure}
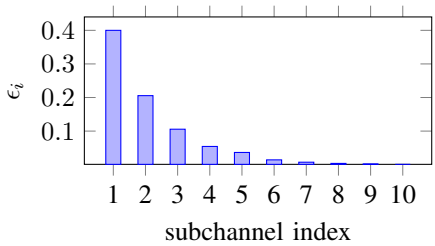

As described in Section~\ref{sec:proposed_soln}, we first train the source mapper and demapper using a series of parallel \glspl{bsc} (multi-level \gls{ber} medium) to learn a binary \gls{jscc} codeword per image.
We consider $N = 10$ \glspl{bsc} with $\epsilon_{1} = 0.4$, $\epsilon_{N} = 0.001$, and
\begin{equation}\label{eq:interface_params}
  \epsilon_{i} = \frac{\epsilon_{1}}{\left( \frac{\epsilon_{1}}{\epsilon_{N}} \right)^{{(i-1)}/{(N-1)}}}, ~ i=2,...,N-1.
\end{equation}
A plot of the \gls{bsc} error probabilities is depicted in Fig.~\ref{fig:subchannel_pe}. The other hyperparameters we used for training are $J=100$ for the number of VIMCO samples and $M=10{,}000$ for the output dimensionality of the source mapper.

While there are various methods for selecting parameters of the multi-level reliability interface, we chose those in (\ref{eq:interface_params}) for the following reasons: For the last reliability level, corresponding to $1{,}000$ bit locations, we set \(\epsilon_{N}=0.001\) to ensure at least one bit error on average, preventing the source mapper from assuming perfect reliability. Given \(\epsilon_{1}\), the intermediate levels were set to maintain a fixed ratio between consecutive parameters. This configuration makes the first reliability level determine the channel's capacity for training the source mapper. 

In the second stage of training, we freeze weights of the source mapper and demapper ($\boldsymbol{\theta}$, $\boldsymbol{\eta}$), and train the channel mapper and demapper weights ($\boldsymbol{\psi}$, $\boldsymbol{\phi}$) using the distortion observed from the source demapper $g_{\boldsymbol{\eta}}$.
In both stages, we use the Adam optimizer \cite{kingma15_adam} with an initial learning rate of $10^{-5}$, which is scheduled to reduce by a factor of $0.8$ if the validation loss does not improve for $2$ consecutive epochs.
We also implement an early stopping policy, terminating each stage if no improvement in the validation loss is observed after $4$ epochs.

\subsection{Numerical results on Split DeepJSCC}
\label{subsec:num_results}

\begin{figure} 
\centering
  \subfloat{%
    \begin{tikzpicture}
        \pgfplotsset{
            legend style={
                font=\fontsize{4}{8}\selectfont,
                at={(1.0,.0)},
                anchor=south east,
            },
            height=0.4\textwidth,
            width=0.5\textwidth,
            xmin=5,
            xmax=20,
            ymin=10,
            ymax=25,
            xtick distance=2,
            ytick distance=2,
            xlabel={$\text{SNR}$ (dB)},
            ylabel={PSNR (dB)},
            grid=both,
            grid style={line width=.1pt, draw=gray!10},
            major grid style={line width=.2pt,draw=gray!50},
            every axis/.append style={
                x label style={
                    font=\fontsize{8}{8}\selectfont,
                    at={(axis description cs:0.5,-0.06)},
                    },
                y label style={
                    font=\fontsize{8}{8}\selectfont,
                    at={(axis description cs:-0.08,0.5)},
                    },
                x tick label style={
                    font=\fontsize{8}{8}\selectfont,
                    /pgf/number format/.cd,
                    fixed,
                    fixed zerofill,
                    precision=0,
                    /tikz/.cd
                    },
                y tick label style={
                    font=\fontsize{8}{8}\selectfont,
                    /pgf/number format/.cd,
                    fixed,
                    fixed zerofill,
                    precision=1,
                    /tikz/.cd
                    },
            }
        }
        \begin{axis}[
        legend columns=2, 
        legend style={
            /tikz/column 2/.style={
                column sep=5pt,
            },
        }
        ]
        \addplot[blue, solid, line width=1.5pt, 
        mark=*, mark options={fill=blue, scale=1.1}] 
        table [x=snr_test, y=10, col sep=comma]
        {rho1_24_split_deepjscc.csv};
        \addlegendentry{\textit{Split DeepJSCC} 
        ($\text{SNR}_{\text{Train}}=10dB$)}
        
        \addplot[green, solid, line width=1.5pt,
        mark=*, mark options={fill=green, scale=1.1}]
        table [x=snr_test, y=15, col sep=comma]
        {rho1_24_split_deepjscc.csv};
        \addlegendentry{\textit{Split DeepJSCC} 
        ($\text{SNR}_{\text{Train}}=15dB$)}
        
        \addplot[red, solid, line width=1.5pt,
        mark=*, mark options={fill=red, scale=1.1}] 
        table [x=snr_test, y=20, col sep=comma]
        {rho1_24_split_deepjscc.csv};
        \addlegendentry{\textit{Split DeepJSCC} 
        ($\text{SNR}_{\text{Train}}=20dB$)}
        
        \addplot[color=black, dashed, line width=1.2pt, mark=*, mark options={fill=black, solid, scale=1.1}] 
        table [x=snr_test, y=2/3_16qam, col sep=comma]
        {rho1_24_digital.csv};
        \addlegendentry{BPG + LDPC 2/3 16QAM}
        
        \addplot[color=magenta, dashed, line width=1.2pt, mark=*, mark options={fill=magenta, solid, scale=1.1}] 
        table [x=snr_test, y=3/4_16qam, col sep=comma]
        {rho1_24_digital.csv};
        \addlegendentry{BPG + LDPC 3/4 16QAM}
        
        \addplot[color=black, dashed, line width=1.2pt, mark=triangle*, mark options={fill=black, solid, scale=1.1}] 
        table [x=snr_test, y=1/2_64qam, col sep=comma]
        {rho1_24_digital.csv};
        \addlegendentry{BPG + LDPC 1/2 64QAM}
        
        \addplot[color=magenta, dashed, line width=1.2pt, mark=triangle*, mark options={fill=magenta, solid, scale=1.1}] 
        table [x=snr_test, y=2/3_64qam, col sep=comma]
        {rho1_24_digital.csv};
        \addlegendentry{BPG + LDPC 2/3 64QAM}
        
        \addplot[color=black, dashed, line width=1.2pt, mark=square*, mark options={fill=black, solid, scale=1.1}] 
        table [x=snr_test, y=3/4_64qam, col sep=comma]
        {rho1_24_digital.csv};
        \addlegendentry{BPG + LDPC 3/4 64QAM}
        \end{axis}
        \end{tikzpicture}
    }
    \\
  \subfloat{%
    \begin{tikzpicture}
        \pgfplotsset{
            legend style={
                font=\fontsize{4}{4}\selectfont,
                at={(1.0,.0)},
                anchor=south east,
            },
            height=0.4\textwidth,
            width=0.5\textwidth,
            xmin=15,
            xmax=30,
            ymin=10,
            ymax=23,
            xtick distance=2,
            ytick distance=2,
            xlabel={$\text{SNR}$ (dB)},
            ylabel={PSNR (dB)},
            grid=both,
            grid style={line width=.1pt, draw=gray!10},
            major grid style={line width=.2pt,draw=gray!50},
            every axis/.append style={
                x label style={
                    font=\fontsize{8}{8}\selectfont,
                    at={(axis description cs:0.5,-0.06)},
                    },
                y label style={
                    font=\fontsize{8}{8}\selectfont,
                    at={(axis description cs:-0.08,0.5)},
                    },
                x tick label style={
                    font=\fontsize{8}{8}\selectfont,
                    /pgf/number format/.cd,
                    fixed,
                    fixed zerofill,
                    precision=0,
                    /tikz/.cd
                    },
                y tick label style={
                    font=\fontsize{8}{8}\selectfont,
                    /pgf/number format/.cd,
                    fixed,
                    fixed zerofill,
                    precision=1,
                    /tikz/.cd
                    },
            }
        }
        \begin{axis}[
        legend columns=2, 
        legend style={
            /tikz/column 2/.style={
                column sep=5pt,
            },
        },
        ]
        \addplot[blue, solid, line width=1.5pt, 
        mark=*, mark options={fill=blue, scale=1.1}] 
        table [x=snr_test, y=22, col sep=comma]
        {rho1_48_split_deepjscc.csv};
        \addlegendentry{\textit{Split DeepJSCC} 
        ($\text{SNR}_{\text{Train}}=22dB$)}
        
        \addplot[green, solid, line width=1.5pt,
        mark=*, mark options={fill=green, scale=1.1}]
        table [x=snr_test, y=25, col sep=comma]
        {rho1_48_split_deepjscc.csv};
        \addlegendentry{\textit{Split DeepJSCC} 
        ($\text{SNR}_{\text{Train}}=25dB$)}
        
        \addplot[red, solid, line width=1.5pt,
        mark=*, mark options={fill=red, scale=1.1}] 
        table [x=snr_test, y=30, col sep=comma]
        {rho1_48_split_deepjscc.csv};
        \addlegendentry{\textit{Split DeepJSCC} 
        ($\text{SNR}_{\text{Train}}=30dB$)}
        
        \addplot[color=black, dashed, line width=1.2pt, 
        mark=*, mark options={fill=black, solid, scale=1.1}] 
        table [x=snr_test, y=2/3_256qam, col sep=comma]
        {rho1_48_digital.csv};
        \addlegendentry{BPG + LDPC 2/3 256QAM}
        
        \addplot[color=magenta, dashed, line width=1.2pt, 
        mark=*, mark options={fill=magenta, solid, scale=1.1}] 
        table [x=snr_test, y=3/4_256qam, col sep=comma]
        {rho1_48_digital.csv};
        \addlegendentry{BPG + LDPC 3/4 256QAM}
        
        \addplot[color=black, dashed, line width=1.2pt, 
        mark=triangle*, mark options={fill=black, solid, scale=1.1}] 
        table [x=snr_test, y=2/3_1024qam, col sep=comma]
        {rho1_48_digital.csv};
        \addlegendentry{BPG + LDPC 2/3 1024QAM}
        
        \addplot[color=magenta, dashed, line width=1.2pt, 
        mark=triangle*, mark options={fill=magenta, solid, scale=1.1}] 
        table [x=snr_test, y=3/4_1024qam, col sep=comma]
        {rho1_48_digital.csv};
        \addlegendentry{BPG + LDPC 3/4 1024QAM}
        
        \end{axis}
        \end{tikzpicture}
        }
\caption{
Comparison between \emph{Split DeepJSCC} and digital baselines:  $\rho = 1/24$ for the top figure and $1/48$ for the bottom figure.\vspace{-0.2cm}}
\label{fig:comparison_digital} 
\end{figure}

We begin by describing the two baselines used for comparison in our study.
The first baseline corresponds to a conventional digital scheme, where the source and channel codes are individually optimized for rate-distortion and \gls{bler} objectives, respectively.
In this setup, we employ BPG, the current state-of-the-art image compression algorithm, for the source code \cite{bellard14_better_portab_graph}. For the channel code, we utilize 5G \gls{ldpc} codes \cite{3gpp_5G_NR_channel_coding}, targeting short blocklengths of $512$ and $768$ bits. 
The second baseline is \gls{deepjscc} \cite{bourtsoulatze19_deep_joint_sourc_chann_codin}, which directly maps the source image to channel symbols without an intermediate bit mapping stage. We remind that \gls{deepjscc} does not conform to the modularized design of current communication networks, where the source and channel are separated by a TCP/IP packet network. However, it still serves as a reference point to gauge the performance of our proposed solution relative to a less constrained setting. To ensure a fair comparison, we use the same architecture for \gls{deepjscc} as the source mapper and demapper depicted in Fig.~\ref{fig:source_encoder_decoder_arch}, with adjustments made to the number of channels in the last convolutional layer of the encoder to maintain consistency in the number of channel uses across all schemes.

Fig.~\ref{fig:comparison_digital} presents the results for $\rho = 1/24$ and $1/48$, corresponding to $K=128$ and $K=64$, respectively. 
The proposed \emph{Split DeepJSCC}, trained with $\text{SNR}_\text{Train}=20\;\text{(dB)}$, demonstrates superior performance compared to the separate source and channel coding baseline across all tested \glspl{snr}. This advantage is more pronounced when the bandwidth is more constrained (i.e., $\rho = 1/48$), as expected, since the benefits of \gls{jscc} are more significant in this scenario.
Additionally, we observe a graceful degradation of image quality with decreasing channel quality, a typical characteristic of \gls{deepjscc} schemes, despite our separate design of the source and channel codecs. 

Another interesting observation is that \emph{Split DeepJSCC}, when trained for higher \gls{snr} values, outperforms models trained for those specific lower \glspl{snr}.
This is different from findings in other \gls{deepjscc} works in the literature, such as \cite{bourtsoulatze19_deep_joint_sourc_chann_codin}, where models trained at a particular \gls{snr} tend to outperform those at other \glspl{snr}.
This phenomenon could be a consequence of the training of the source codec with the multi-level \gls{ber} medium, which generates a codeword with hierarchical order. Different segments of the binary codeword are coded at varying rates, contributing to different aspects of image fidelity. When trained at high \gls{snr}, the channel mapper learns to allocate resources among all the different hierarchies in the codeword, resulting in graceful degradation as \gls{snr} decreases.
This difference in behavior highlights the robustness and adaptability of \emph{Split DeepJSCC} across varying channel conditions.

To obtain results under $\rho = 1/24$ and $1/48$, there is no need to train two separate pairs of source mapper and mapper.
The channel mapper and demapper can be independently trained for each \gls{snr} and $K$, allowing the application layer to design a source code capable of adapting to a wide range of channel conditions.
In this setup, the network operator has the freedom to design its channel code independently. The primary role of the channel mapper and demapper is then to adjust the channel distribution to closely align with what the source codec has been trained for. This separation of responsibilities between the source and channel components enables for greater flexibility and adaptability in the overall system design.

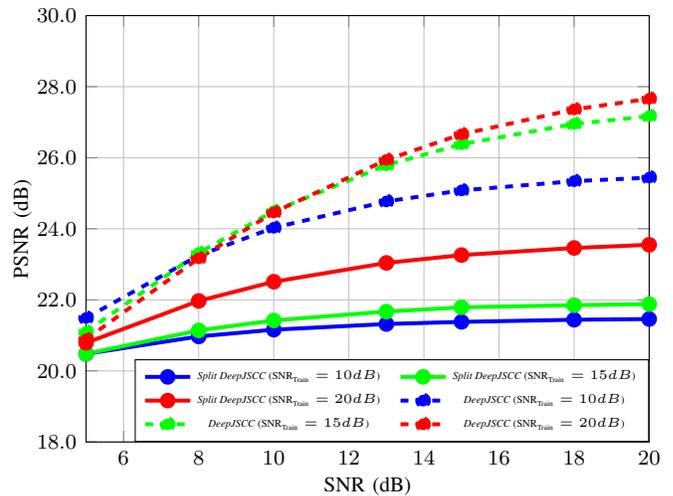
\begin{figure} 
\centering
\begin{tikzpicture}
    \pgfplotsset{
        legend style={
            font=\fontsize{4}{4}\selectfont,
            at={(1.0,.0)},
            anchor=south east,
        },
        height=0.4\textwidth,
        width=0.5\textwidth,
        xmin=5,
        xmax=20,
        ymin=18,
        ymax=30,
        xtick distance=2,
        ytick distance=2,
        xlabel={SNR (dB)},
        ylabel={PSNR (dB)},
        grid=both,
        grid style={line width=.1pt, draw=gray!10},
        major grid style={line width=.2pt,draw=gray!50},
        every axis/.append style={
            x label style={
                font=\fontsize{8}{8}\selectfont,
                at={(axis description cs:0.5,-0.06)},
                },
            y label style={
                font=\fontsize{8}{8}\selectfont,
                at={(axis description cs:-0.08,0.5)},
                },
            x tick label style={
                font=\fontsize{8}{8}\selectfont,
                /pgf/number format/.cd,
                fixed,
                fixed zerofill,
                precision=0,
                /tikz/.cd
                },
            y tick label style={
                font=\fontsize{8}{8}\selectfont,
                /pgf/number format/.cd,
                fixed,
                fixed zerofill,
                precision=1,
                /tikz/.cd
                },
        }
    }
    \begin{axis}[
    legend columns=2, 
    legend style={
        /tikz/column 2/.style={
            column sep=5pt,
        },
    }
    ]
    \addplot[blue, solid, line width=1.5pt, 
    mark=*, mark options={fill=blue, scale=1.1}] 
    table [x=snr_test, y=10, col sep=comma]
    {rho1_24_split_deepjscc.csv};
    \addlegendentry{\textit{Split DeepJSCC} 
    ($\text{SNR}_{\text{Train}}=10dB$)}
    
    \addplot[green, solid, line width=1.5pt,
    mark=*, mark options={fill=green, scale=1.1}]
    table [x=snr_test, y=15, col sep=comma]
    {rho1_24_split_deepjscc.csv};
    \addlegendentry{\textit{Split DeepJSCC} 
    ($\text{SNR}_{\text{Train}}=15dB$)}
    
    \addplot[red, solid, line width=1.5pt,
    mark=*, mark options={fill=red, scale=1.1}] 
    table [x=snr_test, y=20, col sep=comma]
    {rho1_24_split_deepjscc.csv};
    \addlegendentry{\textit{Split DeepJSCC} 
    ($\text{SNR}_{\text{Train}}=20dB$)}
    
    \addplot[blue, dashed, line width=1.5pt, 
    mark=*, mark options={fill=blue, scale=1.1}] 
    table [x=snr_test, y=10, col sep=comma]
    {rho1_24_deepjscc.csv};
    \addlegendentry{\textit{DeepJSCC} 
    ($\text{SNR}_{\text{Train}}=10dB$)}
    
    \addplot[green, dashed, line width=1.5pt,
    mark=*, mark options={fill=green, scale=1.1}]
    table [x=snr_test, y=15, col sep=comma]
    {rho1_24_deepjscc.csv};
    \addlegendentry{\textit{DeepJSCC} 
    ($\text{SNR}_{\text{Train}}=15dB$)}
    
    \addplot[red, dashed, line width=1.5pt,
    mark=*, mark options={fill=red, scale=1.1}] 
    table [x=snr_test, y=20, col sep=comma]
    {rho1_24_deepjscc.csv};
    \addlegendentry{\textit{DeepJSCC} 
    ($\text{SNR}_{\text{Train}}=20dB$)}
    
    \end{axis}
    \end{tikzpicture}
\caption{
Comparison between \emph{Split DeepJSCC} and \gls{deepjscc} schemes under $\rho = 1/24$.
}
\label{fig:comparison_deepjscc} 
\end{figure}

In Fig.~\ref{fig:comparison_deepjscc}, we compare \emph{Split DeepJSCC} with \gls{deepjscc} \cite{bourtsoulatze19_deep_joint_sourc_chann_codin}. Introducing the constraint for the source codec to be separately designed from the channel codec while allowing them to be aware of a shared binary interface, leads to a performance degradation, as expected. However, our approach brings the semantic communication closer to reality. 

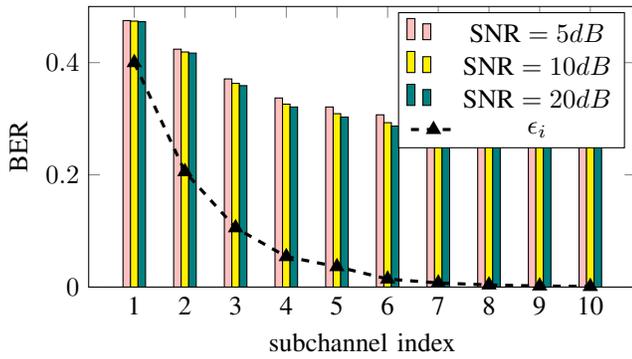
\begin{figure}
    \pgfplotstableread[col sep=comma,]{empirical_ber_snrtrain20.csv}\datatable
        \begin{tikzpicture}
            \begin{axis}[
                width=\linewidth,
                height=0.6\linewidth,
                ymax=0.5,
                ymin=0.0,
                xtick=data,
                xticklabels from table={\datatable}{subchannel},
                ylabel={\gls{ber}},
                xlabel={subchannel index},
                nodes near coords align={vertical}
                ]
                
                \addplot[ybar, ybar legend, bar width=.1cm, fill=pink, bar shift=-.1cm,] table [x expr=\coordindex, y={5}]{\datatable};
                \addlegendentry{$\text{SNR}=5dB$}
                
                \addplot[ybar, ybar legend, bar width=.1cm, fill=yellow] table [x expr=\coordindex, y={10}]{\datatable};
                \addlegendentry{$\text{SNR}=10dB$}
                
                \addplot[ybar, ybar legend, bar width=.1cm, fill=teal, bar shift=+.1cm] table [x expr=\coordindex, y={20}]{\datatable};
                \addlegendentry{$\text{SNR}=20dB$}
                
                \addplot[draw=black, dashed, line width=1.2pt, 
                mark=triangle*, mark options={fill=black, solid, scale=1.1}] 
                table [x expr=\coordindex, y={epsilon}, col sep=comma]
                {bsc_pe.csv};
                \addlegendentry{$\epsilon_i$}
            \end{axis}
        \end{tikzpicture}
    \caption{Empirical \gls{ber} between $\mathbf{u}$ and $\mathbf{v}$, for $\text{SNR}_{\text{Train}}=20dB$and $\rho = 1/24$, compared to the predefined $\epsilon_i$ from the binary interface.\vspace{-0.2cm}}
    \label{fig:empirical_subchannel_ber}
\end{figure}

In Fig.~\ref{fig:empirical_subchannel_ber}, we calculate the empirical \gls{ber} at the output of the channel demapper ($\mathbf{v}$) relative to the input of the channel mapper ($\mathbf{u}$).
We observe that the channel mapper and demapper have learned to achieve \gls{ber} for each subchannel in a pattern that follows the \glspl{ber} assigned to those subchannels in the multi-level interface. This suggests that the channel mapper and demapper have prioritized bits designed to have a greater impact on the distortion.
The substantial difference between the empirical \glspl{ber} in each subchannel and their training values demonstrates not only the resilience of the proposed source codec to errors but also the leniency of the requirement for the network to achieve the expected \gls{ber} in each subchannel.
In practice, although achieving the expected subchannel \gls{ber} is beneficial, a worse \gls{ber} does not lead to a significant degradation in performance at the application level.
This characteristic underscores the practicality of \emph{Split DeepJSCC} as a framework for deploying semantic communication on existing communications infrastructure.

\section{Conclusions}
\label{sec:conclusions}

In this paper, we explored the concept of semantic communications within the constraints imposed by today's communication network infrastructure.
Unlike prior work, we considered the constraint that both the source and network operators should have the freedom to design their coding schemes independently.
To enable this, we proposed a multi-level \gls{ber} binary interface that abstracts the underlying channel, enabling the source to learn a semi-\gls{jscc} codec robust to  channel variations. Once the source code is independently crafted, we train a pair of channel mapper and demapper to map the source bits to channel symbols and vice versa, following the interface BER expectations.
This approach not only allows separate design of the source and channel codes, requiring only the source loss to train the channel mapper, but also enables the channel code to be semantically aware, aligning its performance metric with that of the source.
The results from the proposed solution, which we call \emph{Split DeepJSCC}, demonstrate its ability to fulfill the constraints of current communication infrastructure, making its deployment feasible. Additionally, it achieves many of the benefits observed in prior work that does not adhere to the constraints outlined in this paper, including lower end-to-end distortion and graceful degradation of distortion with declining channel quality. With this contribution, we take a significant step towards realizing semantic communications in modern communication networks.

\bibliographystyle{IEEEtran}
\bibliography{references}

\vfill\pagebreak

\appendix

\subsection{Gradient Analysis for loss function in Eqn~(\ref{eq:cost-per-sample})}\label{sec:appendix_grad}

The targeted loss function is rewritten below,
\begin{equation*}
\mathcal{L}^J(\mathbf{x})\triangleq E_{\hat{\mathbf{u}}^{1:J} \sim p(\hat{\mathbf{u}}^{1:J} | \mathbf{x}, \boldsymbol{\theta})} \left[\log \frac{1}{J}  \sum_{j=1}^J \tilde{p}( \mathbf{x} | \hat{\mathbf{u}}^j, \boldsymbol{\theta},\boldsymbol{\eta}) \right]
\end{equation*}
We first identify the gradient of $\mathcal{L}^J(\mathbf{x})$ with respect to $\boldsymbol{\theta}$,
\begin{equation*}
\begin{split}
    \nabla_{\boldsymbol{\theta}}\mathcal{L}^J(\mathbf{x})&=\nabla_{\boldsymbol{\theta}}\sum_{\hat{\mathbf{u}}^{1:J}} p(\hat{\mathbf{u}}^{1:J} | \mathbf{x}, \boldsymbol{\theta}) \left(\log \frac{1}{J}  \sum_{j=1}^J \tilde{p}( \mathbf{x} | \hat{\mathbf{u}}^j, \boldsymbol{\theta},\boldsymbol{\eta}) \right)\\
    &=\sum_{\hat{\mathbf{u}}^{1:J}} \nabla_{\boldsymbol{\theta}} p(\hat{\mathbf{u}}^{1:J} | \mathbf{x}, \boldsymbol{\theta}) \left(\log \frac{1}{J}  \sum_{j=1}^J \tilde{p}( \mathbf{x} | \hat{\mathbf{u}}^j, \boldsymbol{\theta},\boldsymbol{\eta}) \right)\\
    &+\sum_{\hat{\mathbf{u}}^{1:J}} p(\hat{\mathbf{u}}^{1:J} | \mathbf{x}, \boldsymbol{\theta}) \nabla_{\boldsymbol{\theta}} \left(\log \frac{1}{J}  \sum_{j=1}^J \tilde{p}( \mathbf{x} | \hat{\mathbf{u}}^j, \boldsymbol{\theta},\boldsymbol{\eta}) \right)\\
    &=\sum_{\hat{\mathbf{u}}^{1:J}} \nabla_{\boldsymbol{\theta}} p(\hat{\mathbf{u}}^{1:J} | \mathbf{x}, \boldsymbol{\theta}) \left(\log \frac{1}{J}  \sum_{j=1}^J \tilde{p}( \mathbf{x} | \hat{\mathbf{u}}^j, \boldsymbol{\theta},\boldsymbol{\eta}) \right)\\
    &+\sum_{\hat{\mathbf{u}}^{1:J}} p(\hat{\mathbf{u}}^{1:J} | \mathbf{x}, \boldsymbol{\theta}) \frac{\sum_{j=1}^J \nabla_{\boldsymbol{\theta}} \tilde{p}( \mathbf{x} | \hat{\mathbf{u}}^j, \boldsymbol{\theta},\boldsymbol{\eta})}{\sum_{i=1}^J \tilde{p}( \mathbf{x} | \hat{\mathbf{u}}^i, \boldsymbol{\theta},\boldsymbol{\eta})}.
\end{split}
\end{equation*}

By means of the helper functions $\hat{L}( \hat{\mathbf{u}}^{1:J} )$ and $c_j(\hat{\mathbf{u}}^{1:J})$ defined in Eqn~(\ref{eq:helper_functions}), and since $\nabla_{\boldsymbol{\theta}} \Gamma(\theta)=\Gamma(\theta) \nabla_{\boldsymbol{\theta}} \log \Gamma(\theta)$ for any differentiable function $\Gamma$, we have
\begin{equation*}
\begin{split}
    \nabla_{\boldsymbol{\theta}}\mathcal{L}^J(\mathbf{x})&=\sum_{\hat{\mathbf{u}}^{1:J}} \nabla_{\boldsymbol{\theta}} p(\hat{\mathbf{u}}^{1:J} | \mathbf{x}, \boldsymbol{\theta}) \hat{L}( \hat{\mathbf{u}}^{1:J})\\
    &+\sum_{\hat{\mathbf{u}}^{1:J}} p(\hat{\mathbf{u}}^{1:J} | \mathbf{x}, \boldsymbol{\theta}) \sum_{j=1}^J c_j(\hat{\mathbf{u}}^{1:J}) \log p(\hat{\mathbf{u}}^{1:J} | \mathbf{x}, \boldsymbol{\theta}).
\end{split}
\end{equation*}
Thus,
\begin{equation*}
\begin{split}
    \nabla&_{\boldsymbol{\theta}}\mathcal{L}^J(\mathbf{x})=\sum_{\hat{\mathbf{u}}^{1:J}} p(\hat{\mathbf{u}}^{1:J} | \mathbf{x}, \boldsymbol{\theta}) \nabla_{\boldsymbol{\theta}} \log p(\hat{\mathbf{u}}^{1:J} | \mathbf{x}, \boldsymbol{\theta}) \hat{L}( \hat{\mathbf{u}}^{1:J})\\
    &+\sum_{\hat{\mathbf{u}}^{1:J}} p(\hat{\mathbf{u}}^{1:J} | \mathbf{x}, \boldsymbol{\theta}) \sum_{j=1}^J c_j(\hat{\mathbf{u}}^{1:J})  \nabla_{\boldsymbol{\theta}} \log \tilde{p}( \mathbf{x} | \hat{\mathbf{u}}^j, \boldsymbol{\theta},\boldsymbol{\eta})\\
    &=E_{\hat{\mathbf{u}}^{1:J}} \left[\hat{L}( \hat{\mathbf{u}}^{1:J})\nabla_{\boldsymbol{\theta}} \log p(\hat{\mathbf{u}}^{1:J} | \mathbf{x}, \boldsymbol{\theta}) \right]\\
    &+E_{\hat{\mathbf{u}}^{1:J}}\left[ \sum_{j=1}^J c_j(\hat{\mathbf{u}}^{1:J})  \nabla_{\boldsymbol{\theta}} \log \tilde{p}( \mathbf{x} | \hat{\mathbf{u}}^j, \boldsymbol{\theta},\boldsymbol{\eta})\right],
\end{split}
\end{equation*}
where $\hat{\mathbf{u}}^{1:J}\sim p(\hat{\mathbf{u}}^{1:J} | \mathbf{x}, \boldsymbol{\theta})$.

The gradient of $\mathcal{L}^J(\mathbf{x})$ with respect to $\boldsymbol{\eta}$, takes a similar form. However, since $p(\hat{\mathbf{u}}^{1:J} | \mathbf{x}, \boldsymbol{\theta})$ does not have dependencies on $\eta$, the first term vanishes,
\begin{equation*}
\begin{split}
    \nabla_{\boldsymbol{\eta}}\mathcal{L}^J(\mathbf{x})&=E_{\hat{\mathbf{u}}^{1:J}}\left[ \sum_{j=1}^J c_j(\hat{\mathbf{u}}^{1:J})  \nabla_{\boldsymbol{\eta}} \log \tilde{p}( \mathbf{x} | \hat{\mathbf{u}}^j, \boldsymbol{\theta},\boldsymbol{\eta})\right].
\end{split}
\end{equation*}

\subsection{Proof of Lemma~\ref{lemma:encoder_decoder}}\label{sec:enc_dec_details}

Since the bits in $\mathbf{u}$ are independent Bernoulli random variable, with the means being the outputs of the source mapper $f_{\boldsymbol{\theta}}$, we have
\begin{equation*}
    p( \mathbf{u}_j = u | \mathbf{x},\boldsymbol{\theta})=\begin{cases}
        f_{\boldsymbol{\theta}}( \mathbf{x} )_j& u=1\\
        1-f_{\boldsymbol{\theta}}( \mathbf{x} )_j& u=0.\\
    \end{cases}
\end{equation*}
Here, $f_{\boldsymbol{\theta}}( \mathbf{x} )_j$ denotes the $j$th output of $f_{\boldsymbol{\theta}}( \mathbf{x} )$. Each element in $\hat{\mathbf{u}}$ is a noisy version of the corresponding element in $\mathbf{u}$. In fact, $p( \hat{\mathbf{u}}_j=\mathbf{u}_j)=1-\epsilon_{\lfloor (j-1) N / M \rfloor +1}$ and $p( \hat{\mathbf{u}}_j\neq\mathbf{u}_j)=\epsilon_{\lfloor (j-1)  N/M \rfloor +1}$. Therefore,
\begin{equation*}
\begin{split}
    p( \hat{\mathbf{u}}_j = 1|\mathbf{x},\boldsymbol{\theta})&=p( {\mathbf{u}}_j = 1|\mathbf{x},\boldsymbol{\theta})(1-\epsilon_{\lfloor (j-1) N/M \rfloor +1})\\
    &+p( {\mathbf{u}}_j = 0|\mathbf{x},\boldsymbol{\theta})\epsilon_{\lfloor (j-1) N/M \rfloor +1}\\
    &=f_{\boldsymbol{\theta}}( \mathbf{x} )_j(1-\epsilon_{\lfloor (j-1)N/M \rfloor +1})\\
    &+(1-f_{\boldsymbol{\theta}}( \mathbf{x} )_j)\epsilon_{\lfloor (j-1)N/M \rfloor +1}
\end{split}
\end{equation*}
and $p( \hat{\mathbf{u}}_j = 0|\mathbf{x},\boldsymbol{\theta})=1-p( \hat{\mathbf{u}}_j = 1|\mathbf{x},\boldsymbol{\theta})$. Since $\hat{\mathbf{u}}_j$ is a Bernoulli random variable,
\begin{equation*}
    p( \hat{\mathbf{u}}_j|\mathbf{x},\boldsymbol{\theta})=p( \hat{\mathbf{u}}_j=1|\mathbf{x},\boldsymbol{\theta})^{\hat{\mathbf{u}}_j}(1-p( \hat{\mathbf{u}}_j=1|\mathbf{x},\boldsymbol{\theta}))^{1-\hat{\mathbf{u}}_j}.
\end{equation*}
Further, by design, $p( \hat{\mathbf{u}}|\mathbf{x},\boldsymbol{\theta})=\prod_{j=1}^M p( \hat{\mathbf{u}}_j|\mathbf{x},\boldsymbol{\theta})$. Thus,
\begin{equation*}
\begin{split}
    p&( \hat{\mathbf{u}}|\mathbf{x},\boldsymbol{\theta})=\\
    &\prod_{i = 1}^N \prod_{j = (i - 1)M/N + 1}^{iM/N} (f_{\boldsymbol{\theta}}( \mathbf{x} )_j(1-\epsilon_i)+(1-f_{\boldsymbol{\theta}}( \mathbf{x} )_j)\epsilon_i)^{\hat{\mathbf{u}}_j}\\
    &\hspace{2cm}((1-f_{\boldsymbol{\theta}}( \mathbf{x} )_j)(1-\epsilon_i)+f_{\boldsymbol{\theta}}( \mathbf{x} )_j\epsilon_i)^{1-\hat{\mathbf{u}}_j}\\
    &=\prod_{i = 1}^N \prod_{j = (i - 1)M/N + 1}^{iM/N} \left(f_{\boldsymbol{\theta}}( \mathbf{x} )_j(1-2\epsilon_i)+\epsilon_i\right)^{\hat{\mathbf{u}}_j}\\
    &\hspace{2cm}\left((1-\epsilon_i)+f_{\boldsymbol{\theta}}( \mathbf{x} )_j(2\epsilon_i-1)\right)^{1-\hat{\mathbf{u}}_j}.
\end{split}
\end{equation*}
Finally,
\begin{equation}
\begin{split}
    &\log p( \hat{\mathbf{u}}|\mathbf{x},\boldsymbol{\theta}) \\
    &=\sum_{i = 1}^N \sum_{j = (i - 1)M/N + 1}^{iM/N} \hspace{-0.3cm}{\hat{\mathbf{u}}_j}\log (f_{\boldsymbol{\theta}}( \mathbf{x} )_j(1-2\epsilon_i)+\epsilon_i)\\
    &+\sum_{i = 1}^N \sum_{j = (i - 1)M/N + 1}^{iM/N}\hspace{-0.3cm}({1-\hat{\mathbf{u}}_j})\log((1-\epsilon_i)+f_{\boldsymbol{\theta}}( \mathbf{x} )_j(2\epsilon_i-1))
\end{split}
\end{equation}

As for $\log \tilde{p}( \mathbf{x} | \hat{\mathbf{u}}^j, \boldsymbol{\theta},\boldsymbol{\eta})$, we assume the pixels in $\mathbf{x}$ take a fully factorized Gaussian distribution. Thus, maximizing the Gaussian likelihood (to approximate the best decoder) turns to minimizing the \gls{mse} between the the decoder's reconstruction and the real image, i.e.,
\begin{equation}\label{eq:likelihoodapprox}
\begin{split}
    \log \tilde{p}( \mathbf{x} | \hat{\mathbf{u}}, \boldsymbol{\theta},\boldsymbol{\eta})\approx -|| \mathbf{x} - g_{\boldsymbol{\eta}}( \hat{\mathbf{u}} ) ||_2^2.
\end{split}
\end{equation} 

\subsection{Proof of Lemma~\ref{lemma:ch_encoder_decoder}}\label{sec:ch_enc_dec_details}

The decoded codeword corresponds to $M$ independent Bernoulli random variables with the parameters corresponding to the output nodes of the channel demapper. Thus, $j$-th bit of the decoded codeword is $\mathbf{v}_j=1$, with probability $q_{\boldsymbol{\phi}}( h_{\boldsymbol{\psi}}(\mathbf{u})+\mathbf{n} )_j$, and $\mathbf{v}_j=0$ otherwise, $j\in\{1,\dots,M\}$. Then, by considering the independence of these $M$ random variables, the lemma is concluded.

\end{document}